\documentclass[apj]{emulateapj}
\usepackage{graphicx}
\usepackage{hyperref}
\usepackage{amssymb}
\usepackage{amsmath}
\shorttitle{Low-Metallicity LLSs}
\shortauthors{Cooper et al.}

\providecommand{\zlls}{\ensuremath{z_{\rm LLS}}}
\providecommand{\zqso}{\ensuremath{z_{\rm QSO}}}
\providecommand{\nh}{\ensuremath{n_{\mathrm{H}}}}
\providecommand{\onh}{\ensuremath{\overline{n}_{\mathrm{H}}}}
\providecommand{\NHI}{\ensuremath{N_{\mathrm{H\,I}}}}
\providecommand{\NH}{\ensuremath{N_{\mathrm{H}}}}
\providecommand{\kms}{\ensuremath{\,{\rm km\,s}^{-1}}}
\providecommand{\cm[1]}{\ensuremath{\,{\rm cm}^{#1}}}
\providecommand{\Ang}{\ensuremath{\,\mbox{\AA}}}
\providecommand{\Lya}{\ensuremath{\mathrm{Ly}\alpha}} 
\providecommand{\mh}{\ensuremath{[\mathrm{M}/\mathrm{H}]}}
\providecommand{\xh[1]}{\ensuremath{[\mathrm{#1}/\mathrm{H}]}}
\providecommand{\figm}{\ensuremath{f_\mathrm{IGM}}}

\begin{document}

\title{The Incidence of Low-Metallicity Lyman-Limit Systems at
  $z\approx\!3.5$: \\ implications for the Cold-Flow Hypothesis of Baryonic
  Accretion\altaffilmark{1}}

\author{Thomas J. Cooper,\altaffilmark{2} Robert A. Simcoe,\altaffilmark{1}
  Kathy L. Cooksey,\altaffilmark{2,3} John M. O'Meara,\altaffilmark{4} and Paul Torrey\altaffilmark{2}}
\altaffiltext{1}{This paper includes data gathered with the 6.5 meter Magellan Telescopes located at Las Campanas Observatory, Chile.}
\altaffiltext{2}{MIT-Kavli Center for Astrophysics and Space Research, 77 Massachusetts Avenue, Cambridge, MA 02139, USA}
\altaffiltext{3}{University of Hawai`i at Hilo, 200 West K\=awili Street, Hilo, HI 96720, USA}
\altaffiltext{4}{St. Michael's College, Department of Chemistry and Physics, Colchester, VT 05439, USA}

\begin{abstract}
\noindent Cold accretion is a primary growth mechanism of simulated galaxies,
yet observational evidence of ``cold flows" at redshifts where they
should be most efficient ($z=2$--4) is scarce. In simulations, cold
streams manifest as Lyman-limit absorption systems (LLSs) with
low heavy-element abundances similar to those of the diffuse IGM. Here we
report on an abundance survey of 17 \ion{H}{1}-selected LLSs at
$z=3.2$--4.4 which exhibit no metal absorption in SDSS spectra. Using medium-resolution
spectra obtained at Magellan, we
derive ionization-corrected metallicities (or limits) with a
Markov-Chain Monte Carlo sampling that accounts for the large
uncertainty in $N_{\rm HI}$ measurements typical of LLSs.  The
metal-poor LLS sample overlaps with the IGM in metallicity
and is best described by a model where $71^{+13}_{-11}\%$ are
drawn from the IGM chemical abundance distribution. These represent roughly
half of all LLSs at these redshifts, suggesting that 28--40$\%$ of the
general LLS population at $z\sim3.7$ could trace unprocessed gas. An
ancillary sample of ten LLSs without any {\em a priori} metal-line
selection is best fit with $48^{+14}_{-12}\%$ of metallicities drawn
from the IGM. We compare these results with regions of a moving-mesh
simulation; the simulation finds only half as many baryons in
IGM-metallicity LLSs, and most of these lie beyond the virial radius of the nearest galaxy halo. A
statistically significant fraction of all LLSs have low metallicity
and therefore represent candidates for accreting gas; large-volume
simulations can establish what fraction of these candidates actually
lie near galaxies and the observational prospects for detecting the
presumed hosts in emission.\end{abstract}

\section{Introduction}
Over the past decade, a new theoretical paradigm describing galaxy
evolution and gas accretion has emerged from the synergy between
semi-analytic galaxy formation modeling and high-redshift
observations. In prevailing models of galaxy formation, spiral
galaxies grow largely through a hierarchical merger process, with
relatively quiescent star formation driven by gas accretion onto the
dark matter halo and major mergers initiating periods of rapid
starburst, ultimately resulting in elliptical galaxies with quenched
star formation \citep{1993MNRAS.264..201K,
1998MNRAS.295..319M, 1972ApJ...178..623T,2007ApJ...659..976H}. 
However, recent morphological evidence 
indicates that disk galaxies at high redshifts grow largely through
smooth gas accretion directly onto the stellar
disk \citep{2009ApJ...694L.158B, 2009ApJ...707L...1B},
and mergers may play a less prominent role in their growth
\citep{2013ApJ...771L..35V}. Additionally, the major-merger rate
is too low to explain the number of galaxies at $z\sim2$--3 with a high
star-formation rate \citep[SFR,][]{2008ASPC..396..337J,
2007ApJ...658..763E}, and such mergers do not predict
morphologies seen in galaxies with high SFRs
\citep{2009ApJ...703..785D}. Furthermore, a class of massive
compact spheroidal galaxies with low SFRs is already well established
at $z\sim2$ \citep{2006ApJ...649L..71K,
2008ApJ...677L...5V}, inconsistent with a scenario in which
they are solely a product of major mergers between disk-like galaxies.

Recent simulation work has explored a complementary galaxy
growth mechanism, in which massive galaxies at high redshift are stream-fed large
quantities of gas, and properties of the accreting gas influence the resultant
SFR and morphology \citep[e.g.,][]{2009ApJ...703..785D,2009Natur.457..451D}. Central to this framework is the
existence of ``cold-flow'' accretion---filamentary gas traveling directly
from the intergalactic medium (IGM) onto the star-forming disks of
galaxies, without shock heating at the virial radius. Although
simulations disagree on the exact fraction of gas accreting via cold flows \citep[e.g.,][]{2013MNRAS.429.3353N}, it remains a common feature, fundamental to the growth of early star-forming galaxies.

Observationally, cold flows are expected to manifest as Lyman-limit systems
(LLSs), absorption systems along quasar sightlines with $\tau_{912}\gtrsim2$ ($\log\NHI\gtrsim17.5$), 
in the extended intra-halo medium of galaxies. This environment, often referred to 
as the circumgalactic medium (CGM), forms a regulatory interface 
between galaxies and the IGM and potentially holds a large reservoir of baryons \citep{2013ApJS..204...17W}.

Although the CGM also contains outflowing and recycling gas from the galaxy that
manifests as absorption \citep[e.g.,][]{2014ApJ...796..136B},
the gas metallicity of absorbers serves as a straightforward diagnostic 
to distinguish between accreting baryons and other gas. Outflowing and recycling gas have been
enriched to high metallicities \citep{2010ApJ...717..289S}, often approaching solar, whereas
gas that is being newly introduced to a galaxy is more likely to have low
metallicities consistent with that of the diffuse IGM
\citep{2011MNRAS.418.1796F}. As such, observations establishing
the prevalence of LLSs with pristine elemental abundances would
provide significant evidence supporting cold-flow models of baryonic accretion.

Cold-flow accretion should be most efficient at $2.5\lesssim z\lesssim4.5$, during
the peak of cosmic star formation \citep{2009MNRAS.395..160K}; indeed simulations find
significantly larger covering fractions at $z\approx4$
\citep{2011MNRAS.412L.118F, 2011MNRAS.413L..51K}.
Several groups have reported detections of individual low metallicity
LLSs indicative of cold-flow accretion, but the larger population of
high-$z$ LLSs remains mostly unexplored. \citet{2013ApJ...776L..18C}
report the discovery of a $z=2.44$ LLS with metallicity\footnote{Metallicity is denoted as $[\mbox{X/H}] =\log (N_{\rm X}/N_{\rm H}) - \log(N_{\rm X,\odot}/N_{\rm H,\odot})$, where $N_{\rm X}$ is the column density of an arbitrary atomic species. Often we report \mh\ to indicate ``all metals.''}
$\mh=-2.00\pm0.17$ near a low-luminosity galaxy mixed with 
metal-rich material; \citet{2003A&A...397..851L} discuss a $z=2.92$ LLS
with $\xh{C}=-2.93\pm0.13$; and \citet{2011Sci...334.1245F} present two
LLSs at $z=3.41\ (3.10)$ with upper limits of
$\mh<-4.2\ (-3.8)$. Additionally, two of these absorbers
\citep{2013ApJ...776L..18C, 2011Sci...334.1245F} have
clear deuterium detections with column densities consistent with
primordial abundances, indicating the gas comprising the
absorbers has had little mixing with gas processed by
stars. \citet{2013ApJ...775...78F} construct a composite absorption
spectrum from 20 LLSs at $z\approx\!2.6$--3 selected from a blind QSO
survey and find, for the composite, $\mh\lesssim-1.5$, similar
to the observed metallicities of damped Ly$\alpha$ systems (DLAs, absorbers with $\log \NHI >20.3$).

Studies of the LLS population and CGM are more
extensive at low redshift. \citet{2013ApJ...770..138L} study 28
\ion{H}{1}-selected LLSs at $z\lesssim1$ and find a bimodality in
metallicity, with peaks at $\mh\simeq-1.60$ and $-0.3$. 
Additionally, studies connecting low-redshift metal absorbers with host galaxies find that
the distribution of absorbers is azimuthally and morphologically dependent
in a fashion consistent with a general picture of gas accretion and
galactic winds \citep[e.g.,][]{2012ApJ...760L...7K}, although such studies have
not yet been coupled to metallicity.
The low-metallicity branch of the bimodal distribution is consistent with the notion of cold gas
reservoirs. However, star formation and accretion rates are much lower during
this epoch than at higher redshifts.

In simulations, \citet{2006MNRAS.372..933N} found
the average accretion rate onto galactic halos in $\Lambda$CDM
cosmology goes as $\dot M\propto(1+z)^{2.25}$. 
\citet{2009Natur.457..451D} showed that baryonic-input rates from
cold gas streams in cosmological hydrodynamical simulations follow the
same expression at high redshift. If the \citet{2013ApJ...770..138L} metallicity
bimodality reflects a distinction between accreting gas and enriched outflowing
or recycling gas, then it should
be more pronounced during the peak of cosmic star formation.

There is claimed evidence of cold-flow accretion at low redshift, and
theoretical predictions indicate an increasing frequency with
redshift. However, there are only a handful of high-redshift
observations indicative of cold flows. We seek to determine whether
this is due to an observational shortage or a departure from
expectations from simulations. The Sloan Digital Sky Survey (SDSS) provides a
large, \ion{H}{1}-selected high-redshift sample that can be used to
perform such a study on the population of high-redshift LLSs.

For this paper, we constructed a
survey of high-redshift, \ion{H}{1}-selected LLSs exhibiting low heavy-element 
abundances in SDSS spectra.  In
Section 2, we discuss the selection of sightlines, new observations, and data
processing. Our measurements,
ionization modeling, and metallicity determination of the LLSs are described in Section 3. Section 4 presents
the measured metallicities of the observed LLSs. In Section 5,
we discuss properties of the LLS population, quantify the fraction of
low-metallicity LLSs that are candidates for the observational
signature of cold-flow accretion, and compare with simulations and other observational studies.
Section 6 provides a summary. Throughout, we adopt $\Lambda$CDM cosmology with
cosmological parameters from WMAP9 \citep{2013ApJS..208...19H}: $\Omega_\Lambda=0.72$,
$\Omega_m=0.28$, and $H_0=70\kms\,{\rm Mpc^{-1}}$. 

\section{Data}\label{sec:data} 
We pre-screened a large, \ion{H}{1}-selected sample of LLSs for candidates likely to have low
metallicities and obtained higher resolution spectra of 15 candidates.
This approach maximized information
about the metal-poor end of the LLS distribution, which was not well
understood at the outset. However, it also split the
statistically characterized \ion{H}{1} sample, complicating the
broader interpretation of results.  The pre-screening
proved less efficient than expected in identifying ideal candidates,
roughly halving the parent sample. We
focused our first observations on this metal-poor sub-population, and work exploring the full LLS population is underway.

The parent sample is 194 LLSs with $\zlls\ge3.3$ and
$\NHI\ge17.5\cm{-2}$, compiled by 
\defcitealias{2010ApJ...718..392P}{POW10}\citet[][hereafter POW10]{2010ApJ...718..392P} 
using quasar spectra from the SDSS Data-Release 7 (DR7). They identified systems
by constructing models of the Lyman-series absorption and the Lyman
break and applying them to absorbed quasar continua. Although
they identify more than 194 LLSs, we only consider their ``statistical
sample", comprised of spectra with $\zqso\ge3.6$ and
$\zlls\ge3.3$.

We cross-referenced this list with a \ion{C}{4} $\lambda\lambda1548,1550$ catalog constructed
from the SDSS DR7 quasar spectra \citep{2013ApJ...763...37C} and found
that 152 of the 188 SDSS spectra in our parent sample did not have associated
\ion{C}{4} detections within $\pm500\kms$ of the LLS
redshift. We visually inspected the remaining 152 spectra for typical
metal absorption lines at \zlls. Several spectra had weak \ion{C}{4} doublets below thresholds of the
\ion{C}{4} survey or \ion{C}{4} obstructed by interlopers, and many spectra did not have definitive \ion{C}{4}
but had absorption from \ion{C}{2} $\lambda1334$ or other low-ionization
species. Ultimately, the LLS sample was roughly halved by a metal-line inspection, with definite or probable metal absorption lines associated with 100 of the LLSs, and no corresponding metals seen for 96 of them.

The metal-poor LLS sample presented here was subjected to an additional declination
cut for observation at the Magellan telescopes, since they are situated at a latitude of -29$^{\circ}$ and the SDSS footprint is primarily in the northern sky. Excluding quasars with $\delta>+21^{\circ}$ (corresponding to a transit airmass of $\approx1.5$) leaves 28 sightlines, 15 of which we observed in this initial survey.

\subsection{MagE Spectra}

We obtained higher resolution spectra along 15 quasar sightlines (Table \ref{llstab}) selected as described above
using the Magellan Echellete Spectrograph \citep[MagE,][]{2008SPIE.7014E.169M} on the 6.5-m Magellan Clay telescope. MagE covers optical wavelengths from $3100\Ang$ to $1\,\micron$. At $\zlls=3.3$, the 912\Ang\ Lyman break is redshifted to 3900\Ang. With an $0.85\arcsec$ slit, MagE has a resolution $\mathcal{R}=4950$ (or full width at half maximum ${\rm FWHM} =60.7\kms$). Observations were done on UT 17/19 March 2013 and UT 05 May 2013 with typical seeing of 0.6\arcsec--0.8\arcsec.

Data were reduced using the MASE pipeline \citep{2009PASP..121.1409B},
using GJ 620.1 B/HIP 80300 as a standard for calibration.  MASE is an
IDL software package designed for reducing MagE data and performs
the full extraction and calibration process. We manually
construct a cubic-spline fit to the continuum of each spectrum.

\begin{figure}[t]
\plotone{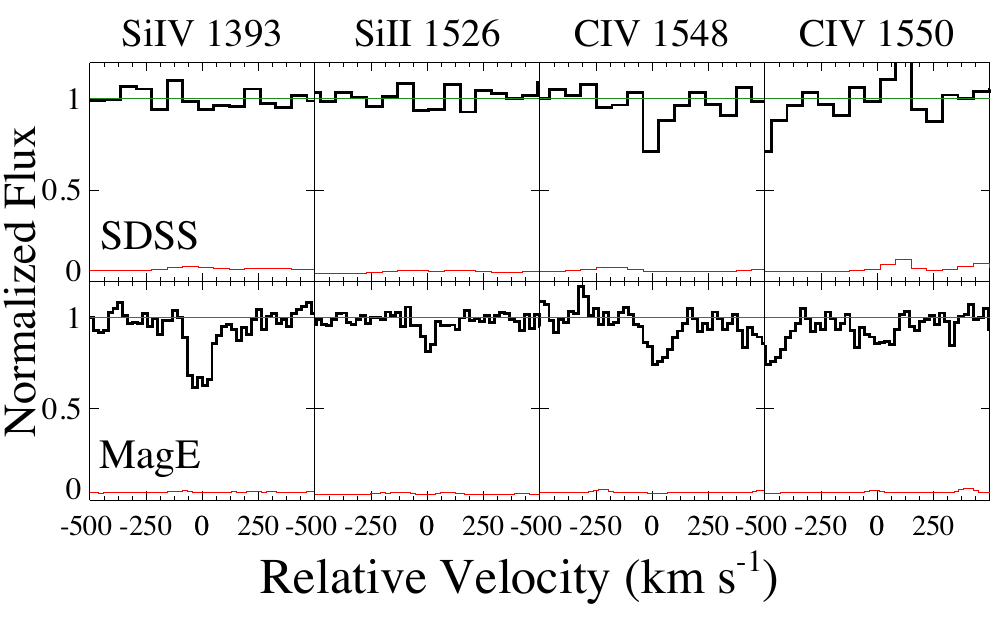}
\caption{Comparison of normalized spectra (black) of J083832 from SDSS and MagE. The red line is the 1$\sigma$ error; the green line is the continuum (unity). The spectrum cutouts are centered around \zlls\ ($v=0\kms$). The
  \ion{Si}{4} $\lambda1393$ and \ion{Si}{2}
  $\lambda1526$ lines and \ion{C}{4} doublet are all evident in the MagE spectrum, but are not seen
  in the SDSS spectrum. \ion{Si}{4} $\lambda1402$ is unavailable in both
  spectra due to a strong interloping absorber.}\label{compare}
\end{figure}

Figure \ref{compare} shows portions of the SDSS and MagE spectra of
J083832 around several of the LLS metal lines for comparison.
In this example there is no statistically significant metal absorption from the LLS
in the SDSS spectrum (${\rm FWHM} \approx 150\kms$) but the MagE spectrum has clear absorption lines.

\subsection{Higher Resolution and Infrared Spectra}

Several of the LLSs have no metal absorption in
their corresponding MagE spectra. We observed one such object, 
J124957, with the Magellan Inamori Kyocera Echellette spectrograph 
\citep[MIKE,][]{2003SPIE.4841.1694B} at the same telescope on UT 20 March 2013
using a $1\arcsec$ slit. MIKE is a double echelle spectrograph. The
blue arm covers $3200\Ang$ to $5100\Ang$, and the red arm
covers $4900\Ang$ to $1\,\micron$. With a $1\arcsec$ slit, the
blue\slash red arms have resolutions of 28,000\slash 22,000 (${\rm FWHM} =
10.7\kms$\slash 13.6\kms). All metal lines used in the current survey are in the red
portion of the spectrum. MIKE data were reduced using \texttt{MIKE
Redux},\footnote{\url{http://web.mit.edu/~burles/www/}.} a series of IDL
tools that encompass all calibrations and extractions.

We also make use of a medium-resolution infrared spectrum
of the same object taken with the Folded-port InfraRed Echellette \citep[FIRE,][]{2008SPIE.7014E..0US},
on the 6.5-m Magellan Baade telescope, observed as part of a different survey.
FIRE has a bandpass covering $0.8\micron$--$2.5\micron$, at a resolution of 
$\mathcal{R}=6000$  (${\rm  FWHM}=50\kms$). For data acquisition and reduction details
see \citet{2012ApJ...761..112M}.

The primary motivation for the MIKE observation was the possibility of identifying deuterium 
absorption associated with the \ion{H}{1}, which can indicate that gas is unprocessed \citep[e.g.,][]{2013ApJ...776L..18C}. Unfortunately, the \ion{H}{1} absorption from the LLS along this sightline proved too broad to distinguish deuterium absorption. However, the higher resolution MIKE spectrum enables more sensitive column-density measurements,
and the FIRE spectrum allows us to measure several ions not covered by the optical instruments.

\subsection{Metal-Blind Sample}
We are interested in how our sample of metal-poor LLSs compares to the
global LLS population. As a comparison sample, we studied spectra from the blind LLS
survey of \cite{2013ApJ...775...78F}, also conducted with MagE. Since their
sample is at slightly lower redshifts than ours, we selected the ten
highest-redshift absorbers from their survey (excluding several 
DLAs and absorbers close to the QSO redshift) to
achieve the best redshift overlap with our metal-poor sample. This method of choosing
objects also avoids introducing any selection bias with regards to metallicity. The ten LLSs we examined have a median redshift of \zlls=3.04. For the remainder
of the paper, we refer to this dataset as the ``metal-blind
sample," and to our observations as the ``metal-poor sample." 
We applied identical analysis techniques to reduced spectra in both samples.

\begin{figure*}[t]
\plotone{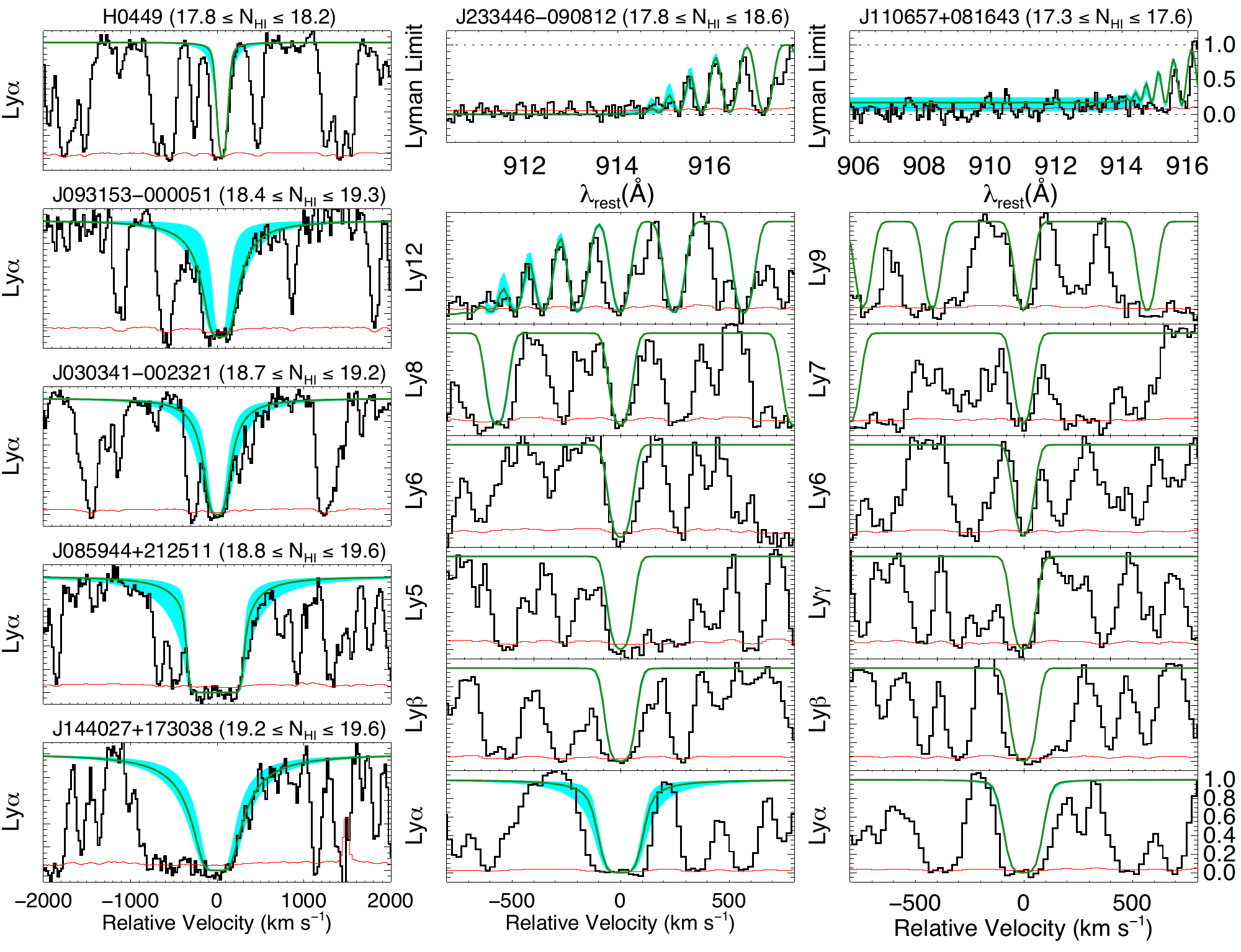}
\caption{Velocity plots of example \ion{H}{1} profile fits. The black curves are the normalized spectra, and the red curves are the $1\sigma$ error on the flux. Green lines are representative fits, and the cyan shadings fill the regions between the Voigt profiles corresponding to the smallest and largest plausible \ion{H}{1} column densities.
{\bf Left:} Several Ly$\alpha$ absorption profiles, showing how Ly$\alpha$ fitting contributes in different regimes of \NHI.  The bottom profile has damping wings that tightly constrain \ion{H}{1}. The  figures above this show how Ly$\alpha$ can place an upper limit on \NHI\ by requiring model profiles to not over-absorb.
{\bf Center:} Several Lyman series transitions drawn from the same system. For this absorber, Higher order Lyman series transitions (e.g., Ly6 $\lambda930$, Ly8 $\lambda923$, Ly12 $\lambda917$) are used to measure the Doppler $b$ parameter, and lines nearing the Lyman limit are fit to measure \NHI\ (top panel). The saturation at the Lyman limit also places a lower limit on \NHI.
{\bf Right:} Example of a comparatively weak absorption system. The absorption is not fully saturated at the Lyman limit, allowing an \NHI\ estimation. The absorption profile does not vary within the range of \NHI\ allowed by the flux seen at the Lyman limit (top panel).}
\label{nhi}
\end{figure*}

\section{Analysis}\label{analysis}

We inspected the Lyman series absorption in each of the MagE spectra
to confirm LLS redshifts found in \citetalias{2010ApJ...718..392P} using SDSS spectra. We found a
difference in redshift of $\lesssim\!0.01$ for all but one of the
systems. The outlier J085944 has a weaker absorber ($\log \NHI<17.5$)
at the redshift determined by \citetalias{2010ApJ...718..392P}. The redshift of the LLS
($\zlls=3.263$) is smaller by $\approx\! 0.2$ and is the lowest redshift
LLS in our study.  Since we did not check for metal absorption in
the SDSS spectrum at the correct redshift, this system could have
biased our sample. However, the absorber serendipitously met the
target selection criteria discussed in Section \ref{sec:data}, so we included
it in our analysis. Additionally, two of the quasar spectra had a
second, lower-redshift LLS close enough in redshift to target system
that enough of the Lyman series transitions were available to measure
the \ion{H}{1} column density. These two LLSs also meet the selection criteria (no metals seen in the corresponding SDSS spectrum)
and were included in the analysis.

\subsection{\ion{H}{1} Column Density}

\NHI\ is notoriously difficult to measure in the LLS regime; the
\Lya\ curve-of-growth is flat and higher order transitions, including
the Lyman break, are saturated by construction. We found that for
the purpose of calculating metallicities, ionization modeling
techniques can marginalize over a wide range of \ion{H}{1} column
densities, at least within the LLS range, as we discuss in Section
\ref{valid_sec}.

Rather than attempting to find \ion{H}{1} explicitly using Voigt
profiles, we determined a range of viable column densities for each
system, listed in Table \ref{llstab}. We used modified versions of IDL
software from the
\texttt{XIDL}\footnote{\url{http://www.ucolick.org/~xavier/IDL/}}
library in conjunction with the Voigt profile fitting packages
\texttt{VPFIT} and
\texttt{RDGEN}\footnote{\url{http://www.ast.cam.ac.uk/~rfc/vpfit.html}}.

For each system, the plausible range of \NHI\ is determined by fitting
various different aspects of the \ion{H}{1} absorption signature
(Figure \ref{nhi}). We estimated the Doppler $b$ parameter, a measure of
the width of the Voigt profile, using higher order Lyman series
transitions. For weaker systems where the Lyman limit is not fully
saturated, we were able to constrain \NHI\ using the flux at the Lyman
limit and/or fitting higher order Lyman series lines. For the
strongest absorption systems, we fit the weak damping wings on the
Ly$\alpha$ profile. For systems of middling strength, the saturation
of the Lyman limit coupled with the non-existence of Ly$\alpha$
damping wings restrict the range of possible \NHI.

The typical \NHI\ uncertainty (defined as max(\NHI)-min(\NHI) for a given system) for both samples is 0.7 dex. This median total deviation is akin to an error bar of $\pm$0.35 dex. The best constrained system had a total deviation of 0.3 dex, while the least constrained had a deviation of 1.7 dex, although the maximum is an outlier with the next largest being 1.3 dex. These errors are incorporated into our metallicity uncertainty as bounds on a flat prior distribution of allowed \NHI, defining the range where we explore possible values for our solutions.

\subsection{Metal Column Densities}

We measure column densities for ionic species using the apparent
optical depth method \citep[AODM,][]{1991ApJ...379..245S}.  For each
absorber, we integrated over a fixed velocity width in order to maintain
consistency between different species\slash lines (rounded to the nearest
pixel). For ions with multiple available lines, we performed an
error-weighted average of the detections.

For each line where there was a non-detection, we calculated a 3-$\sigma$
upper limit to the column density corresponding to the error in the
spectrum over $\pm1$ resolution element using 10,0000 Monte-Carlo
realizations. For each iteration, we added to the flux a value drawn from
a Gaussian distribution with width equal to the error at that pixel,
then measured the column density of the resulting mock profile,
discarding realizations where the column density was negative.
If an equal amount of flux were scattered above and below the continuum level, 
this would result in a positive column density because the relationship
between flux and apparent optical depth is exponential 
\citep[see][]{2005AJ....130.2418F}. This produced a distribution of the
largest possible column densities consistent with the
observed lack of absorption. We adopted the column
density larger than $99.7\%$ of other trials as the 3-$\sigma$ upper limit. To mitigate the
effects of poor continuum fits resembling low-column density absorption, we set
the flux to unity and repeated the process if less than 50\% of the
trials produce a positive column density.

Similar to the \ion{H}{1} analysis, we dismissed all metal absorption
lines contaminated by interloping absorbers as well as lines
obscured by large amounts of noise. 
The \ion{Si}{2} $\lambda1260$
transition, which is a powerful diagnostic for low-metallicity
absorbers due to its large oscillator strength, was unavailable for
ten of the 17 systems due to confusion with the Ly$\alpha$ forest.

All AOD column density measurements assumed the
absorption profiles reside on the linear portion of the curve-of-growth
and are unsaturated.  While this assumption is expected to hold
for all of the lines in our metal-poor sample, 
there is no absorption strength cut for the metal-blind sample so we need to test for saturation.
In low- and medium-resolution spectra, absorption profiles can be saturated without clearly exhibiting zero flux,
since the instrument blurs the absorption profile. In Table \ref{cdenstab} we list the measured column densities, and below we discuss the identification of saturated lines.

\subsubsection{Testing for Saturation}

The AODM provides a test for saturation through
comparison of the the AOD profiles for different
transitions of the same species \citep{1991ApJ...379..245S}. However, this is insufficient when a species only has a 
single, potentially saturated line. To test for saturation in such species, we performed multi-component Voigt profile fits to see if we recover column densities similar to those measured with AODM. Since the velocity structure of absorbers is typically not well resolved in our spectra, we used Monte Carlo methods to explore the parameter space for each line. For each line, we constructed 200 models to use as input to \texttt{VPFIT}, each having 3--5 components (uniformly selected) placed by splitting the absorber into bins and uniformly selecting redshifts within these bins, such that components extend over the entire absorption profile. Following measurements of \ion{C}{4} Doppler $b$ parameters \citep{1996ApJ...467L...5R},
components have Doppler parameters drawn from a Gaussian distribution with $\bar{b}=12\kms$ and
$\sigma_b=5\kms$, constrained to be above 6\kms, with $b$ held
fixed during fitting. We allowed \texttt{VPFIT} to
remove components and slightly modify redshifts.

\begin{deluxetable}{llllll}
\tablecaption{Details on specific absorbers\label{cdenstab}}
\tablehead{
\colhead{Ion}&\colhead{$\lambda_{\textrm{rest}}$}&\colhead{$\log N_\mathrm{AODM}$}&\colhead{$\log$ $N_\mathrm{adopted}\tablenotemark{a}$}}
\startdata
\cutinhead{Metal-Poor Sample}
\cutinhead{J080853-070940\hspace{.5 cm}$z$ = 3.545}
\ion{C}{4}&1548&$13.38\pm0.13$&$13.39\pm0.11$\\
\ion{C}{4}&1550&$13.42\pm0.21$&\nodata\\
\ion{O}{1}&1302&$<13.65$&$<13.65$\\
\ion{Si}{2}&1304&$<13.31$&$<13.30$\\
\ion{Si}{2}&1526&$<13.45$&\nodata\\
\ion{Si}{4}&1393&$12.93\pm0.12$&$12.93\pm0.12$\\
\ion{Si}{4}&1402&$<13.36$&\nodata\\
\cutinhead{J083832+200142\hspace{.5 cm}$z$ = 3.47595}
\ion{Al}{2}&1670&$12.73\pm0.04$&$12.73\pm0.04$\\
\ion{C}{2}&1334&$14.29\pm0.02$&$>14.29$\\
\ion{C}{4}&1548&$13.65\pm0.05$&$13.70\pm0.04$\\
\ion{C}{4}&1550&$13.85\pm0.07$&\nodata\\
\ion{Si}{2}&1526&$13.49\pm0.10$&$13.49\pm0.10$\\
\ion{Si}{4}&1393&$13.52\pm0.03$&$13.52\pm0.03$
\enddata
\tablecomments{This table will be published in its entirety in the electronic edition; a portion is shown here as an example.}
\end{deluxetable}

The intent of this exercise was not to determine specific models for
the unresolved velocity structure of the absorbers but rather to gauge whether
there is saturation by seeing if the models with column densities
larger than measured by the AODM provided reasonable fits to the
spectra. In evaluating the output, we first removed all fits where the
structure was reduced to a single component (which tends to produce
poor fits with unrealistically large column densities, since the
Doppler parameters were fixed) and all fits where the $\chi^2$-fit
statistic output by \texttt{VPFIT} was more than $2\sigma$ above the mean (these are typically trials 
where the profile is essentially a single
component fit with several negligible components). The remaining
models provide a distribution of potential column densities for the
absorbing ion.

We considered an ion to be saturated if it met two requirements: (i)
less than 5\% of the trials had column densities less than that
obtained via AODM and (ii) the median column density of the modeled distribution
exceeded $N_\mathrm{AODM}+{\rm max}(3\sigma_\mathrm{AODM},0.2\,{\rm 
  dex})$.  The second criterion prevented lines with Monte Carlo trials resulting in very
 precise, narrow output distributions only slightly larger
 than the AODM measurement from being falsely classified as saturated.
When we determined that an ion was saturated,
we adopted the AODM measurement as a lower limit to its column density.

We found this approach agreed with both our expectations for which lines are
saturated based on appearance and AODM testing for saturation. For very weak absorption profiles this technique did not
reliably produce meaningful results, but comparing AOD profiles when there are multiple ions shows that such lines are clearly unsaturated, as expected. In our metal-blind sample, we found only one saturated line, a \ion{C}{2} $\lambda 1334$ line that was misattributed to QSO \ion{H}{1} self-absorption in the SDSS spectrum, both due to its location on the QSO Ly$\alpha$ emission peak and the lack of corroborating lines. There are numerous saturated lines in the metal-blind sample, as indicated in Table \ref{cdenstab}.

\begin{figure}[t]
\epsscale{1.15}
\plottwo{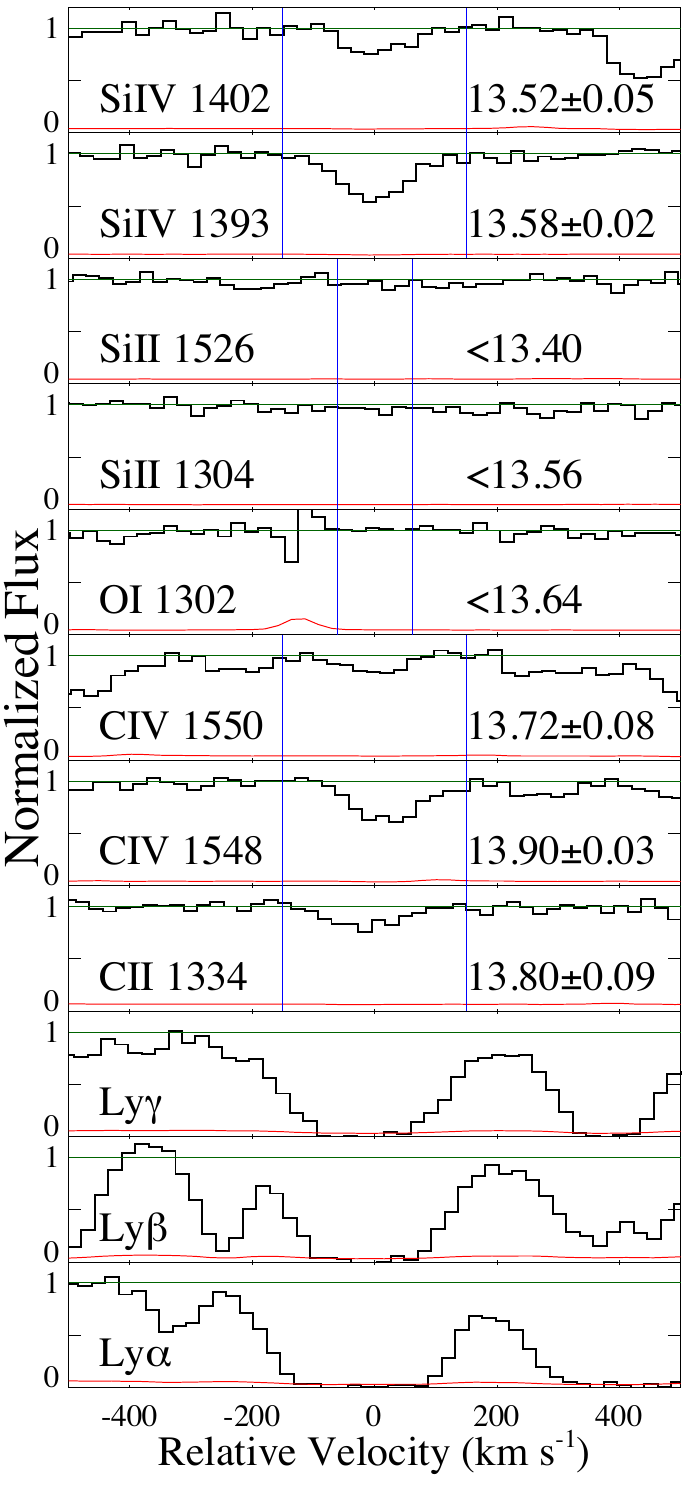}{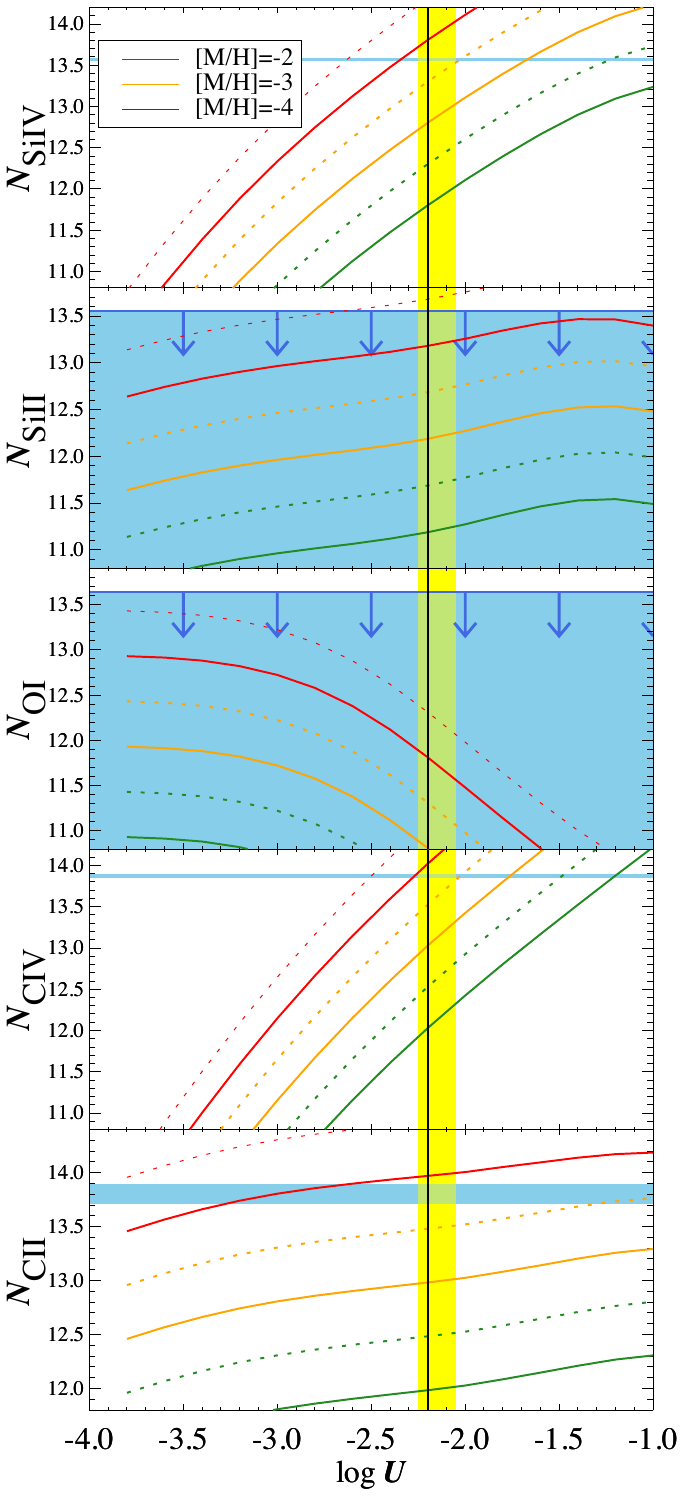}
\caption{Absorption profiles and example ionization model for J123525.
 {\bf Left}: The normalized spectrum (black) of J123525 at the position of
various metal lines relative to \zlls\ ($v=0\kms$), with 1$\sigma$ uncertainty on the flux in red. The green line indicates the continuum (unity).
The blue lines indicate the range over which the optical depth was
integrated to determine the column density. Where no absorption was detected, 
the blue lines are instead $\pm$ one
resolution element from the central redshift, indicating the range
over which the 3$\sigma$ upper limit was measured.
 {\bf Right}: Example ionization model. For each ion, the solid
and dashed curves show the column density as a function of the
ionization parameter for different metallicities. The blue shaded regions are
1$\sigma$ intervals around the column density detections or, for non-detections, the region below the upper limit. The yellow shaded region is the range of $\log U$ found to have a consistent solution.
The black line is the ionization parameter corresponding to the best-fit
solution.}
\label{J123525}
\end{figure}

\subsection{Ionization Modeling and Metallicity Determination}

The primary interest of our study is the metallicity,
\mh, of the absorbers. However,
this requires knowledge of elemental abundance ratios, rather than the
information on specific ions that we measure. Rather than assuming
ionization conditions to convert between ionic and atomic column
densities, we used the software package \texttt{Cloudy} \citep[version 13.02, last described by,][]{2013RMxAA..49..137F} to solve for
the ionization and metallicity simultaneously. With \texttt{Cloudy}, we modeled
the ionization conditions of the absorbers over a range of
metallicities, obtained ionic column densities for each model, and
determined which models best matched the observed column densities
using Monte Carlo simulations. Ionization conditions are typically
described by the ionization parameter $U$, a proxy for hydrogen density \nh\ defined
as:
\begin{align}
U=\frac{\Phi}{n_Hc}\label{eq:u}
\end{align}
where $c$ is the speed of light and $\nh=n_{\rm H\,I}+n_{\rm H\,II}+n_{\rm H_2}$. The flux of \ion{H}{1}-ionizing photons, $\Phi$, is given by
\begin{align}
\Phi=4\pi \int_{\nu_{\rm LL}}^{\infty}\frac{J_\nu}{h\nu}d\nu\label{eq:phi}
\end{align}
where $J_\nu$ is the specific intensity of the incident radiation and $\nu_{\rm LL}$ is the frequency corresponding to 1 Ryd.

\begin{figure}[t]
\epsscale{1.15}
\plottwo{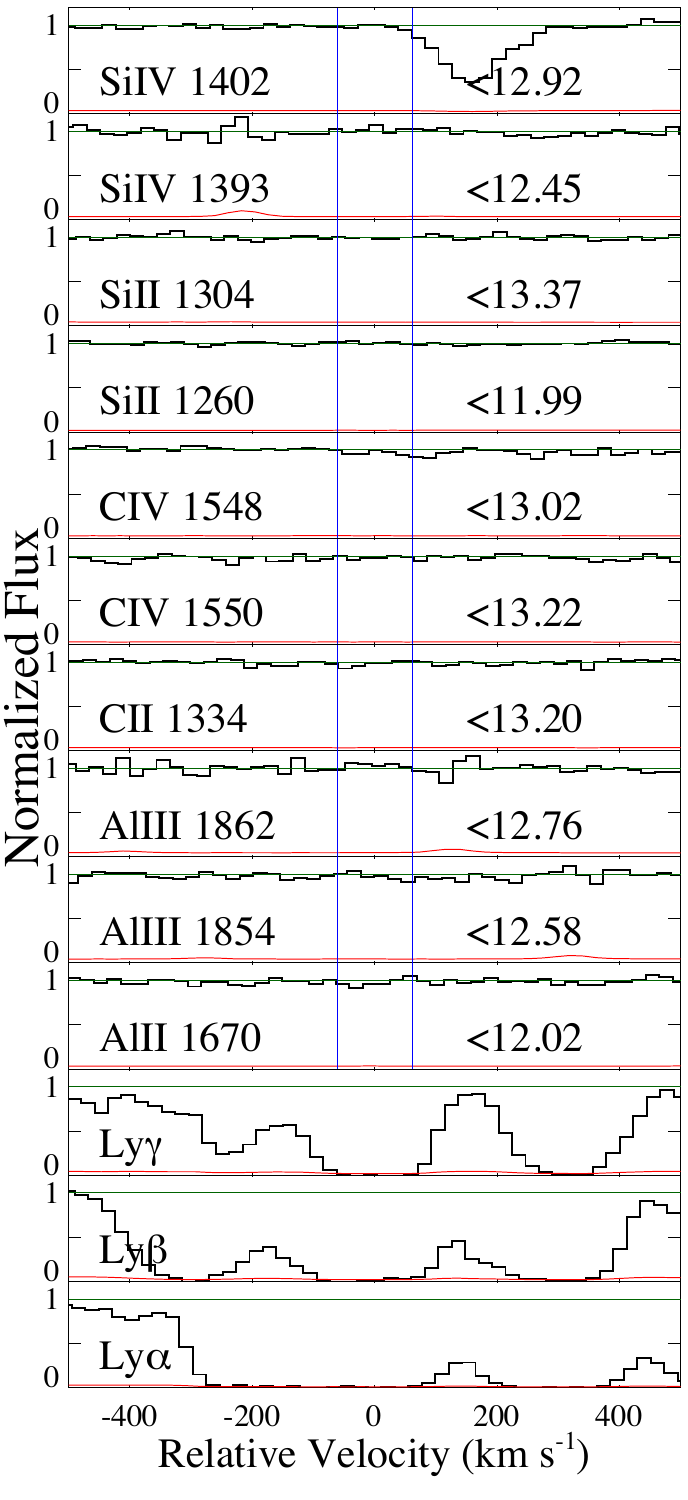}{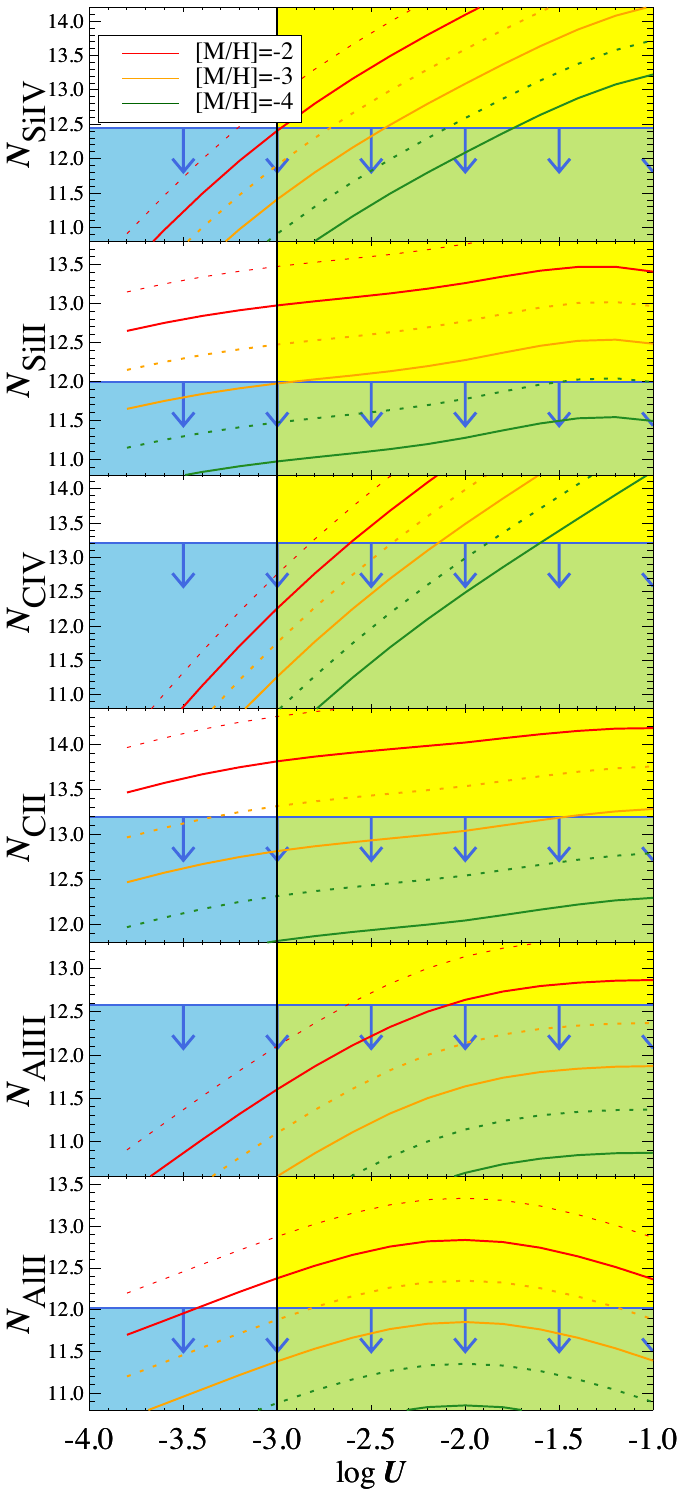}
\caption{Same as Figure \ref{J123525} but for J124957. 
 {\bf Left}: The normalized spectrum of J124957 at the position where it's LLS's various metal lines would be. Since no absorption is
detected for any of these lines we only obtain upper limits.
 {\bf Right}: Without any metal column density measurements to
constrain the ionization parameter, we assume $\log U\ge-3$ and measure the metallicity
upper limit at $\log U=-3$. For each ion, the limiting metallicity is
at the upper-left corner of the overlapping shaded regions;
the lowest of these gives the upper limit metallicity for the LLS. For
this system, the strongest constraint at $\log U=-3$ comes from \ion{Si}{2}
1260.}
\label{J124957}
\end{figure}

Specifically, we used \texttt{Cloudy} to calculate the column densities of
different species as a function of $\log U$ over a grid of
\ion{H}{1} column densities, redshifts, and metallicities. The models were
generated assuming a plane-parallel geometry of a
uniform and isothermal layer of photoionized gas, with the shape of
the ionizing radiation spectrum derived from a combination of the cosmic microwave background (CMB)
and the cosmic ultraviolet background (UVB) spectrum from
\citet[][CUBA software]{2012ApJ...746..125H}. The CMB is much weaker than the
UVB at all relevant wavelengths and omitting it
caused no appreciable change. We adopted 
a solar relative abundance pattern for the \texttt{Cloudy} models.

The UVB spectrum includes emission
from galaxies and QSOs, as well as a sawtooth absorption pattern due
to the \ion{He}{2} Lyman series \citep{2009ApJ...693L.100M}.  Although observations
of the hydrogen ionizing flux disagree with the normalization of this spectrum \citep{2013MNRAS.436.1023B,2008ApJ...688...85F}, \texttt{Cloudy} models adjust the normalization according to the input value of $U$, so this is inconsequential. Efforts to adjust the shape of the spectrum have found that best-fit models typically require fairly small modifications that ultimately translate to a difference in metallicity of $\lesssim0.2$ dex \citep{2015MNRAS.446...18C}, although there are outliers that are best fit with larger modifications to the shape. Since our modeling is based on a small number of ions, we keep the number of fit parameters to a minimum to avoid overfitting and  do not let the shape of the spectrum vary in our models. Additionally, this ionizing background (without adjustment) was used in most of the works we compare to in Section \ref{sec:compobs}, so any uncertainties in the shape of the spectrum should minimally affect comparison with other observations.

For each LLS, we interpolated on this grid to the absorber redshift, producing for each ionic metal species the
column density as a function of \mh, $\log U$, and \NHI. In Figures \ref{J123525} and \ref{J124957} we sketch the ionization modeling technique for LLSs with and without metal-line absorption, respectively. These examples assume a value of \NHI\ intermediate to the allowed range, so one dimension is missing from this schematic representation of the modeling process. 

\subsubsection{Dependence of Results on \NHI} \label{sec:nhidep}

\begin{figure}
\plotone{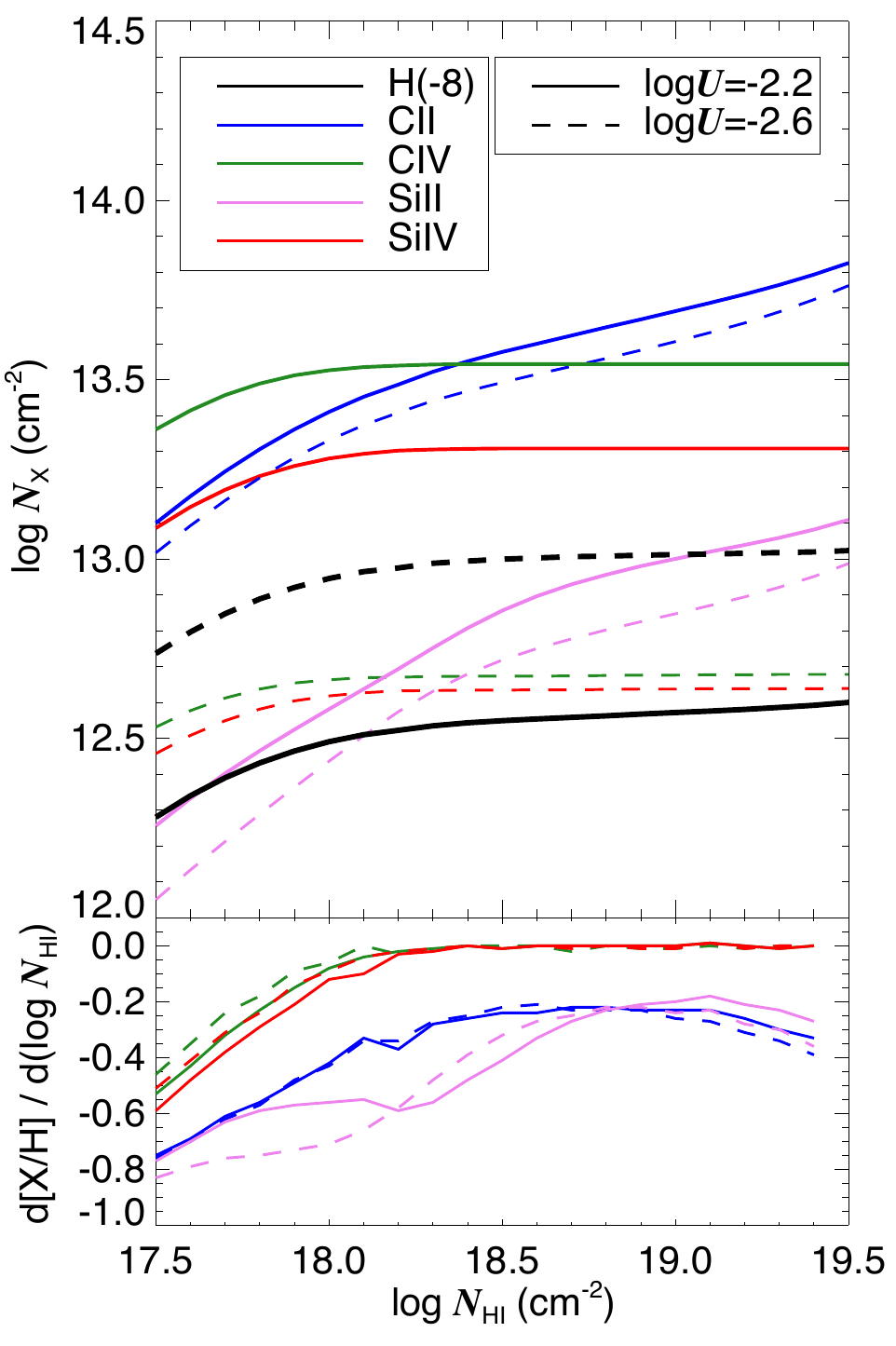}
\caption{{\bf Top:} Model column densities for several ions, assuming $\log U=-2.2$ (solid curves) and $\log U=-2.6$ (dashed curves) with fixed metallicity $\mh=-2.5$. The black curve is the total hydrogen column density $N_H$, scaled by $10^{-8}$. The intermediate ions (\ion{C}{4} and \ion{Si}{4}), along with \NH\ are quite flat with \NHI. The low ions are more correlated with \NHI\, although the range of column densities they span is still about one-tenth of that considered in \NHI. {\bf Bottom:} Slope of the metallicity as measured by a single ion and assuming a fixed value of $\log U$. This can be used to estimate the uncertainty in \mh\ introduced by the uncertainty in \NHI, although this approach overestimates \mh\ uncertainty since it does not consider overlapping constrains from multiple ions.}
\label{nvnhi}
\end{figure}

Although many systems require us to marginalize over a fairly broad range in \NHI, ionization modeling within the bounds we consider is surprisingly insensitive to the particular \ion{H}{1} column density. Just as Figures \ref{J123525} and \ref{J124957} are projections onto an assumed value of \NHI, in Figure \ref{nvnhi} we project along two different values of $\log U$ (at fixed $\mh=-2.5$), to clarify how several properties vary with \NHI.

In the top panel, we plot the model column density as a function of \NHI\ for several different ions, as well as the total hydrogen column density \NH, scaled to fit on the same plot. For a given value of $\log U$, \NH\ does not scale very rapidly with \NHI; over the two orders-of-magnitude of \NHI\ shown, \NH\ only changes by about 0.3 dex. The column density curves for \ion{C}{4} and \ion{Si}{4} are comparably flat. The variation in the low-ionization metals with \NHI\ is more appreciable, but is still one order-of-magnitude smaller than the variation in \NHI. We do not show the scaling with \mh, although it is as expected--the metal column densities increase by an order-of-magnitude when \mh\ is increased by one. Since \NH\ and most of the metal column density curves are relatively flat functions of \NHI, the model parameters $U$ and \mh\ corresponding to the model that matches measured ionic column densities from an LLS do not vary strongly with \NHI.

Comparing the model column density curves for the two different ionization parameters plotted, we see that the low ions are not strongly dependent on $\log U$. \ion{C}{4} and \ion{Si}{4} column densities depend much more strongly on how ionized the gas is, as can also be seen in the right-hand plots of Figures \ref{J123525} and \ref{J124957}.

We estimate from the \texttt{Cloudy} output how a metallicity measurement (at fixed $\log U$) based on a single ion varies with \NHI. For ion x corresponding to atom X, with measured column density $N_x$ and model ionization fraction $f_x=N_x/N_X$, we can write the metallicity as
\begin{align*}
\xh{X}&=\log N_X/\NH-\log(N_X/\NH)_{\odot}\\
&=\log\frac{f_\mathrm{HI}N_x}{f_x\NHI}-\log(N_X/\NH)_{\odot}\\
\end{align*}
where $f_\mathrm{HI}$ is the \ion{H}{1} ionization fraction in the corresponding model.
Noting that the measured column density is a constant, we can differentiate with respect to $\log\NHI$:
\[\frac{\partial\mh}{\partial\log\NHI}=\frac{\partial\log f_\mathrm{HI}}{\partial\log\NHI}-\frac{\partial\log f_x}{\partial\log\NHI}-1\]
The derivative depends only on ionization model output, and is shown in the bottom panel of Figure \ref{nvnhi} for several ions.

We first note that all of the slopes are negative, indicating that metallicity decreases with \NHI\ as expected. The \ion{Si}{4} and \ion{C}{4} slopes approach zero above $\NHI\sim18$, indicating that modeled metallicities should be very robust in this range of column densities---given $N_\mathrm{CIV}$ or  $N_\mathrm{SiIV}$ and $\log U$, the metallicity is independent of \NHI. \ion{Si}{2} and \ion{C}{2} have larger derivatives with \NHI, as expected from the column density curves, but weaker dependence on $\log U$.

This suggests that uncertainty of \NHI\ is manageable. Treating the metallicity slope as a proxy for the model uncertainty in the resulting metallicity and integrating the \ion{C}{2} derivative from \NHI=17.5 to 18.5, where the model uncertainty is worst, results in only a $\sim~0.5$ dex uncertainty in metallicity. In practice, the uncertainties are significantly less since models are fit using multiple ions.

\subsubsection{Matching Ionization Models to LLSs}

To compare how well different ionization models fit the data, we define the likelihood function 
\begin{displaymath} \mathcal{L}(\mh,\NHI,\log U)=\Pi_i\ell_i(N_i)
\end{displaymath} 
where $\ell_i(N_i)$ is the likelihood for each ion measured given $N_i$, the model column density obtained by interpolating the grid to the corresponding values of \mh, \NHI\ and $\log U$. For detections, $\ell_i$ is taken to be a Gaussian with a mean and standard error given by the AODM measurements. For upper- and lower-limits, we let $\ell_i$ be unity if the model column density is within the range allowed by the limits, and $\ell_i$ decays as a Gaussian with $\sigma=0.05$ beyond the allowed range. In practice, we used $\log\mathcal{L}$ to avoid computational instabilities resulting from small likelihoods.

\begin{figure*}[t]
\plottwo{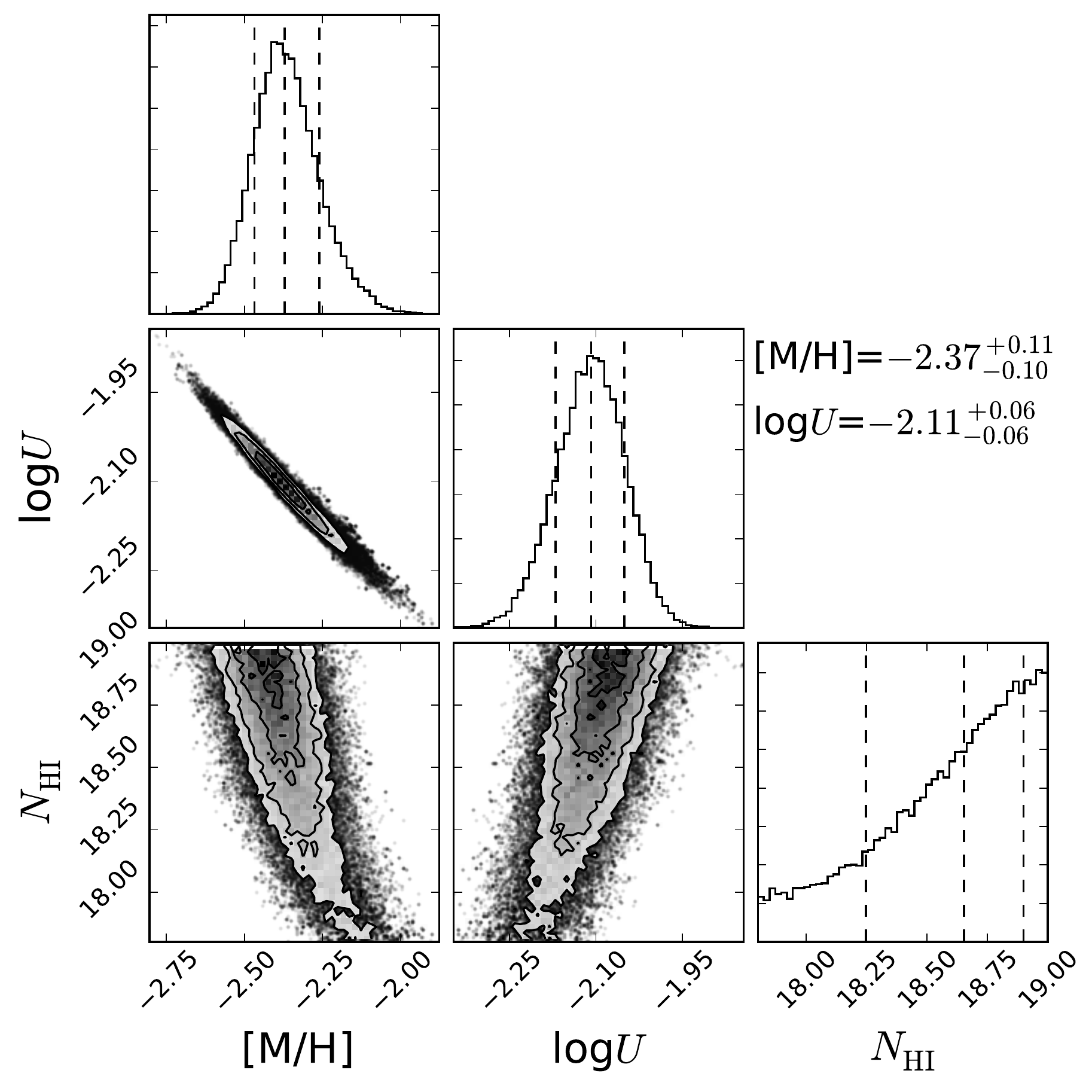}{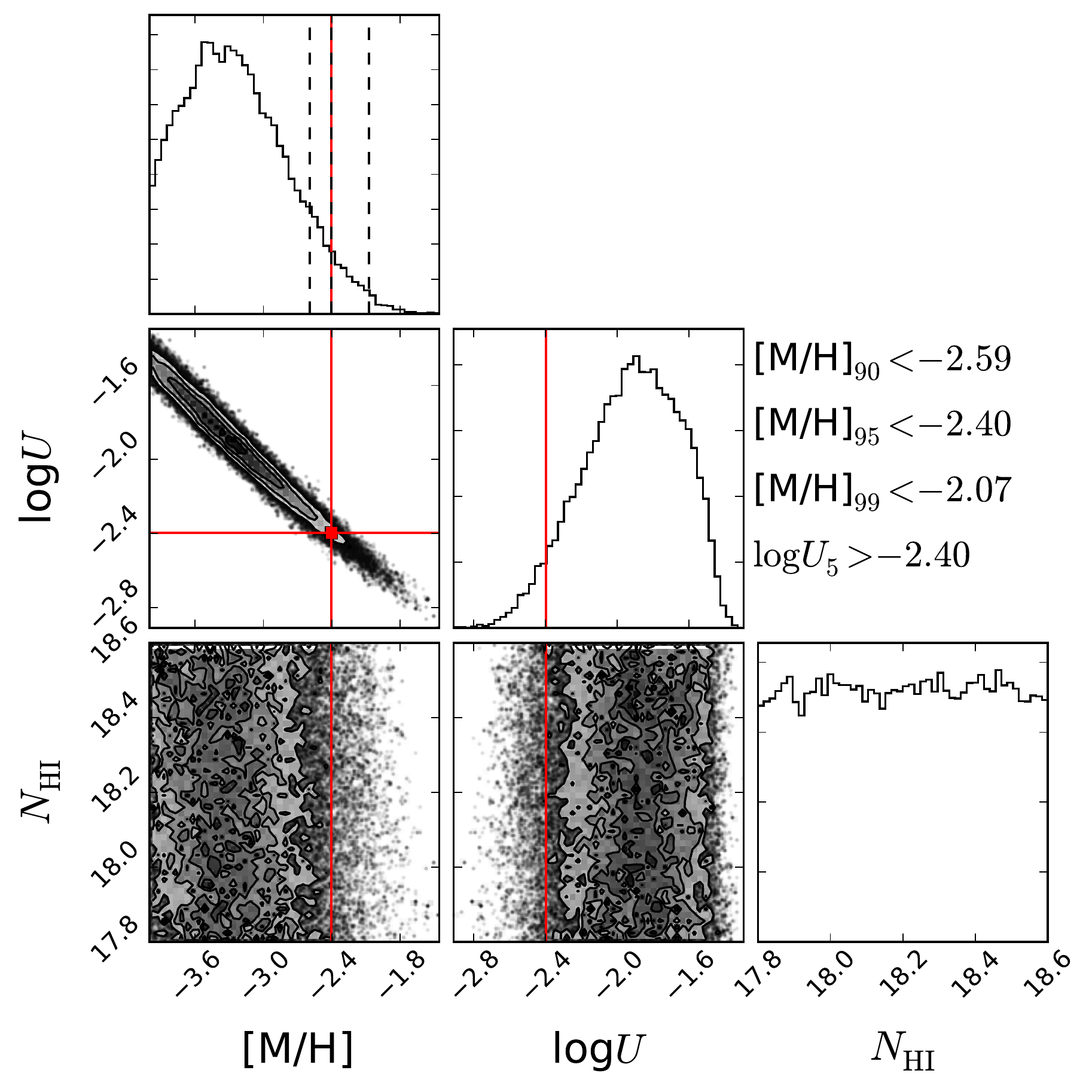}
\plottwo{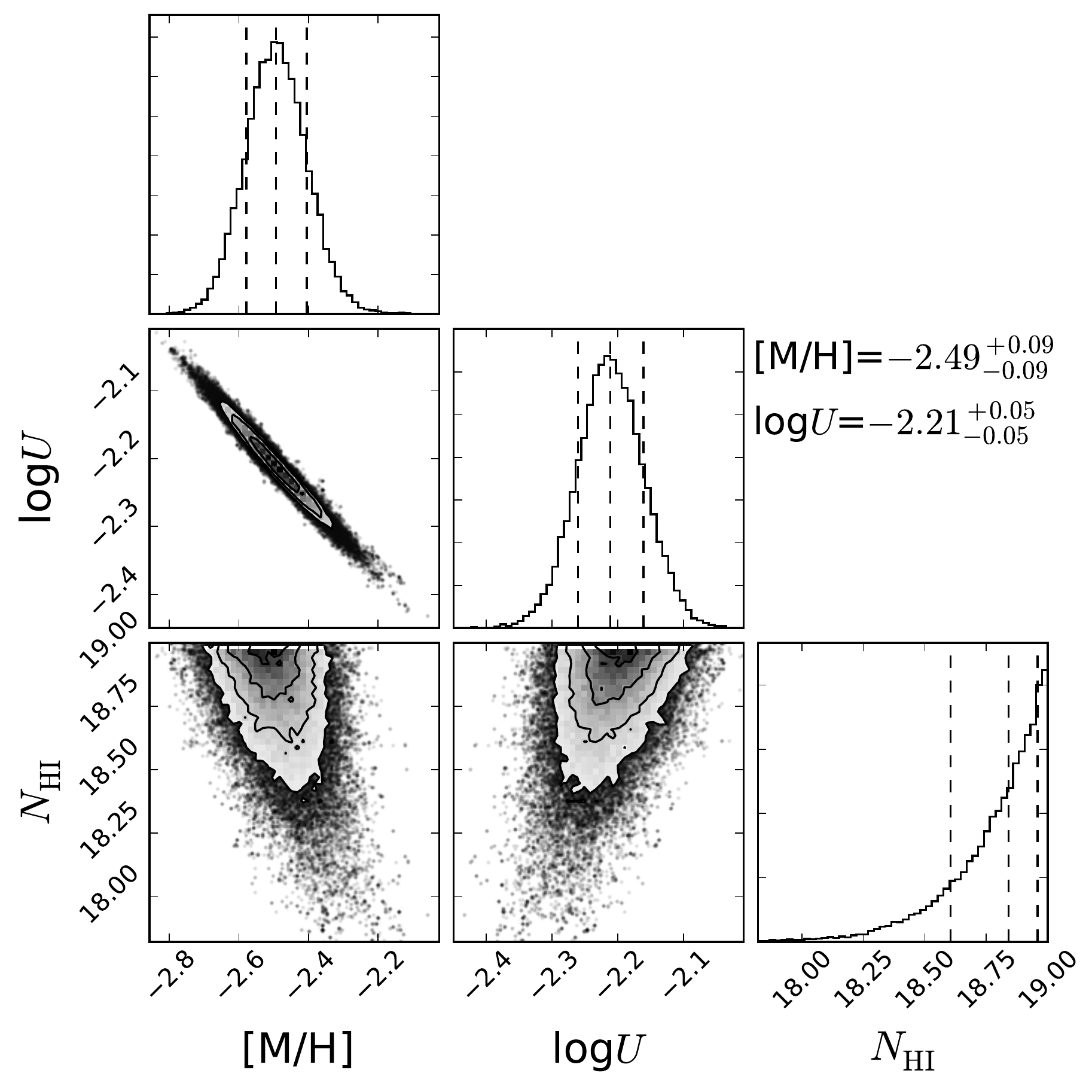}{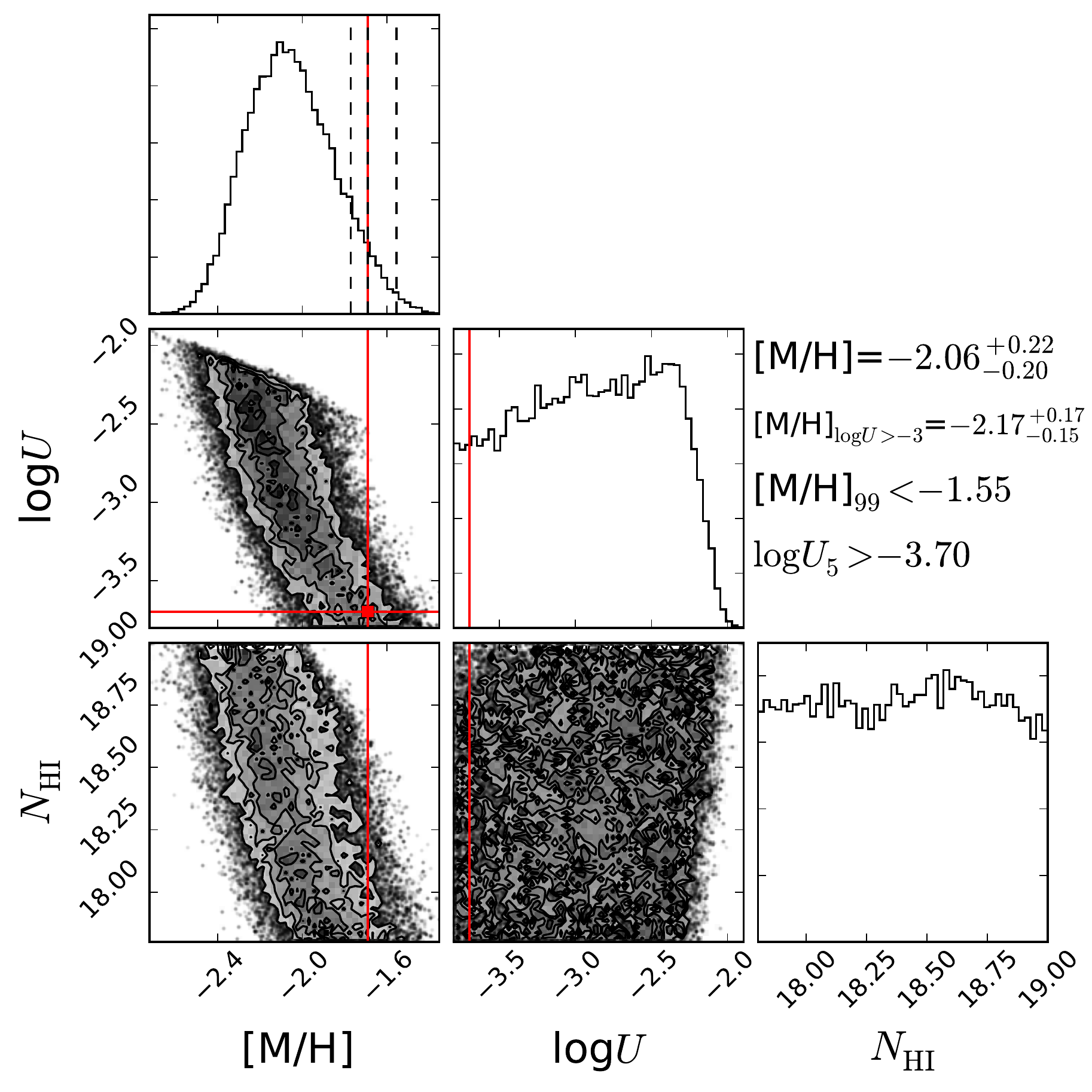}
\caption{{\bf Top Left} Posterior distribution for ionization modeling of J123525. Histograms of \mh, $\log U$, and \NHI are along the diagonal, with dashed lines showing the 16th, 50th, and 84th percentiles. The other three plots show the positions of the walkers at each step, outlining correlations between the three variables. \mh\ and $\log U$ are not strongly dependent on \NHI, but \mh\ and $\log U$ are tightly correlated. The width in the \mh--\NHI\ profile reflects the range of acceptable ionization parameters. The bias in \NHI\ results from moderately larger likelihoods for the metal column densities at larger \NHI\ values for this system, but does not appreciably influence the results. {\bf Top Right} Same as top left, for J080853. This system gives a Type 2 upper limit: there are upper limit constraints on the metallicity (i.e., lower limit constraints on the ionization parameter), but the walkers do not converge to a solution. The dashed lines correspond to the 90th, 95th, and 99th percentiles. The red lines are the 95th percentile in \mh\ and the 5th percentile in $\log U$ (since $U$ and \mh\ are inversely related). {\bf Bottom Left} J123525 (same as top left), but only half the \ion{C}{4} and \ion{Si}{4} column density, to test the effect of interloping gas that is coincident but unaffiliated with the LLS. The posterior is largely similar to that using the measured column densities. {\bf Bottom Right} J123525 again, but with \ion{C}{4} and \ion{Si}{4} treated as upper limits. In this scenario, the analysis yields a metallicity upper limit. Figures formatted with the Python module \texttt{Triangle} \citep{dan_foreman_mackey_2014_11020}}
\label{mcmc}
\end{figure*}

Motivated by the implementation of \citet{2015MNRAS.446...18C}, we used the Python module \texttt{emcee} \citep{2013PASP..125..306F} to perform a Markov Chain Monte Carlo (MCMC) sampling of the parameter space. This approach allows for calculation of the posterior probabilities of the metallicity and ionization parameter while marginalizing over the possible values of \NHI. We used flat priors with $\log U$,$\mh\in[-4,-1]$, and \NHI\ within the viable range determined for each LLS. We ran 1000 iterations of 100 walkers sampling the parameter space, discarding the first 100 iterations as `burn-in'  to allow the walkers to explore the full posterior distribution and to remove the signature of the walkers' initial conditions. We constructed the posterior distribution from the remaining 900 iterations. An example posterior distribution (for J123525) is shown in the top left of Figure \ref{mcmc}.

Although many systems required marginalization over a wide range in \NHI, often larger than one order-of-magnitude, we found that the ionic column densities and posterior probabilities are not strongly dependent on \NHI, consistent with expectations from the discussion in Section\,\ref{sec:nhidep}.

\subsubsection{Metallicity Upper Limits}

For LLSs where the absorption line data were insufficient to constrain the solution, we derived upper limits to the metallicity. The way an upper limit was found depends on whether or not there were any metal-line detections. We note that as $\log U$ increases, \nh\ decreases and the gas becomes more highly ionized,
resulting in column densities for the ions we measure to correspond to lower metallicities at
larger ionization parameters (Figures \ref{J123525} and \ref{J124957}, and Figure \ref{mcmc} left-middle panel). Hence, lower values of $\log U$ correspond to more conservative upper bounds on metallicity.

For absorbers with only non-detections, we found the limiting metallicity at $\log U=-3$, a
conservative estimate for high-redshift systems consistent with other works \citep{2011Sci...334.1245F}. We also only consider the smallest allowed value of
\NHI, since this corresponds to the most conservative upper limit.
The upper limit is the highest metallicity for which all model column
densities (at the smallest \NHI\ allowed) were less than the measured limits at $\log U=-3$. We refer to these as ``Type 1" upper limits. Figure \ref{J124957} is an example of a Type 1 upper limit. The
strongest column density constraint for this example came from \ion{Si}{2}
1260 due to its large oscillator strength, although the Ly$\alpha$
forest made it unavailable for many of our systems. 

For absorbers with some detections but not enough to fully constrain the posterior distribution in \mh-$\log U$ space,
the metallicity upper limit is derived from the posterior distribution. We take the upper limit to be 95th percentile of the posterior metallicity distribution. We refer to these as ``Type 2" upper limits. J080853 is presented as an example in the top right of Figure \ref{mcmc}. Note that the posterior distribution for $\log U$ does not extend below $-3$---if it did, then we would reclassify this system as a Type 1 upper limit. 

Type 2 limits generally result in lower metallicity limits than Type 1, since measured metal column densities are able to constrain the ionization parameter to a larger limiting value. Since the posterior distribution for these systems includes the largest ionization parameters and lowest metallicities, the exact value taken as the limit depends on the range of \mh\ allowed by our priors, but in practice our prior distribution was realistic enough that large modifications are not physically motivated, and small changes to the priors have negligible effect on the result.

\subsubsection{Applicability of Single Cloud Models}\label{valid_sec}

Our analysis treated each absorption system as a single cloud comprised
of gas with minimal phase structure, which is insufficient to
capture the full structure and conditions of the
CGM. \cite{2015ApJ...802...10C} constructed mock absorption lines
through the CGM of a simulated $z=0.54$ dwarf galaxy, investigating
the kinematics and phase structure of the gas. For low-ionization
metals (e.g., \ion{Mg}{2} $\lambda\lambda2796,2803$) as well as \ion{H}{1}, absorption along their
simulated sightlines was generally dominated by a narrow, single-phase
cloud that could be readily modeled with typical ionization correction
techniques. However, absorption from high-ionization gas (e.g., \ion{O}{6} $\lambda\lambda1031,1036$, \ion{C}{4}) often came from more extended structures with varying gas
properties, as well as gas unassociated with the \ion{H}{1}
but with a coincident line-of-sight velocity.

This suggests that the influence of \ion{C}{4} and \ion{Si}{4} on our
results needs to be examined, to gauge whether or not they adversely affect 
our analysis. If we assume that an appreciable fraction of
the high-ionization metal absorption along a sightline were due to gas
more extended than or disjoint from the \ion{H}{1}, then the measured high-ionization metal
column densities would be upper limits to the associated \ion{H}{1} column densities.

To assess how this would influence our ionization modeling, we use the LLS along the sightline to J123525 (Figure \ref{J123525}) as an example. If we treat the \ion{C}{4} and \ion{Si}{4} as upper limits, then the only ion with a measured column density is \ion{C}{2}. Performing the ionization modeling under this assumption we found a posterior distribution essentially inverse that of a Type 2 upper limit (Figure \ref{mcmc}, bottom right): the metallicity is constrained, but the ionization parameter is not. This could be treated as a Type 1 upper limit, which would have $\mh\le-1.5$, much larger than the metallicity measured treating \ion{C}{4} and \ion{Si}{4} as detections. The explanation for the large change can be seen in Figure \ref{nvnhi}. As discussed in Section \ref{sec:nhidep}, model \ion{C}{2} column density is moderately dependent on \NHI, so without any other lines constraining the result, taking the metallicity limit at the smallest viable value of \NHI=17.8 and $\log U$=--3 gives a high-metallicity upper limit. When all metal lines are treated as detections, the metallicity is much lower at the same \NHI\ because other detections limit the possible solutions. 

However, since there is a \ion{C}{2} detection, treating this as a Type 1 upper limit ignores critical information. An actual Type 1 upper limit posterior distribution would have both \mh\ and $\log U$ extending to $\mh=-4$, the low metallicity cutoff of our priors. The \ion{C}{2} detection constrains the metallicities for a given value of $\log U$ and restricts the maximum value of $\log U$. The posterior distribution for this example gives $\mh=-2.06$, fairly close to nominal metallicity found for this system in spite of the low values of the ionization parameter that are included in the posterior. Restricting the posterior to $\log U>$-3, the modeling gives \mh=--2.17. Hence, treating \ion{C}{4} and \ion{Si}{4} as upper limits for this LLS change the posterior distribution for $\log U$, but \mh\ only changes by several tenths of a dex.

Even if some of the high-ionization metals are from interloping gas, ionization modeling still predicts a sizable column density from the LLSs. If, for J123525, instead of treating the \ion{C}{4} and \ion{Si}{4} as upper limits, we take half of the observed column density to be from the LLS, the MCMC posterior distribution (Figure \ref{mcmc}, bottom left) is very similar to that found using the actual measured values, with a median metallicity of $\mh=-2.49$, slightly lower than the metallicity we measure without altering the measured column densities.

The results found for this example generally extend to other systems with both high- and low-ionization detections: treating high ions as detections leads to the inclusion of low $\log U$ values in the posterior distribution, but ultimately only influences the metallicity by less than 0.5 dex. 

\begin{figure}[t]
\plotone{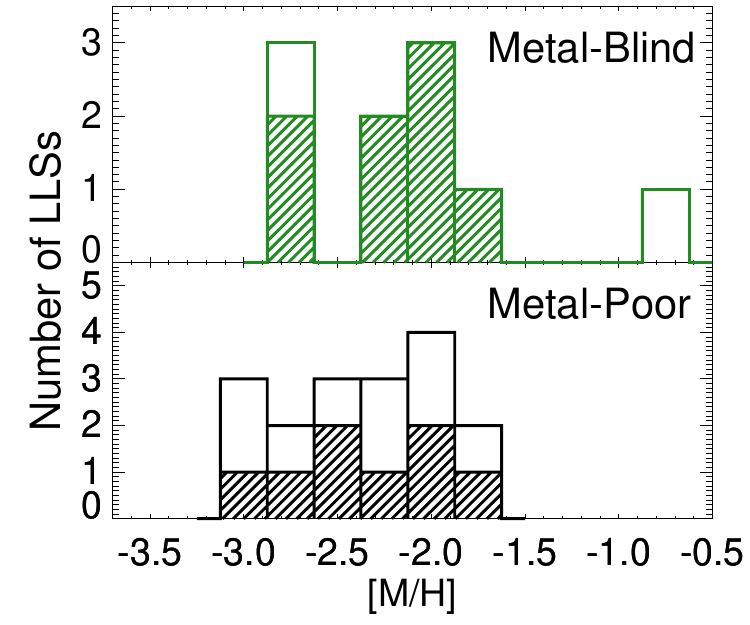}
\caption{Metallicities with a bin size of 0.25\,dex
  centered on half-integer values. The shaded region consists of
  detections, whereas the unshaded portions correspond to upper
  limits, except for lower limit in the metal-blind sample above \mh=-1. Although the distribution of the detections is not markedly different, the metal-poor sample contains far more metallicity upper limits.}
\label{m_histo}
\end{figure}

There are several LLSs for which we detect only high-ionization metals, most of which are Type 1 or 2 upper limits. Since \ion{C}{2} and \ion{Si}{2} non-detections tend to place strong constrains on metallicities, treating \ion{C}{4} and \ion{Si}{4} detections as upper limits for these systems tends to produce metallicity upper limits within 0.5 dex of the measured limits. There is one outlier, J115321, a Type 2 upper limit that would become a Type 1 upper limit with a limiting metallicity that is 1 dex larger. There are two LLSs in the metal-blind sample (J234466 \& J1025909) with metallicity solutions, but only \ion{C}{4} and \ion{Si}{4} detections. We found these systems have posterior distributions highly constrained by \ion{Si}{2} and \ion{C}{2} non-detections, such that treating the high-ionization detections as upper limits produces Type 1 metallicity upper limits that are only $\sim$0.5 dex larger than the measured metallicities.

\section{Results}

In Table \ref{llstab} we list the properties of the quasars and LLSs comprising our metal-poor and metal-blind samples. Of the 17 metal-poor LLSs, nine have metallicity upper limits---six Type 1 limits derived from five LLSs with no metal detections along with one LLS with only \ion{C}{4}, and three Type 2 limits from systems without information enough information for the MCMC walkers to converge. In addition to having generally higher metallicities, the metal-blind survey has only one upper limit (Type 2). The metal-blind sample also has one metallicity lower limit, derived from several ions with saturated absorption profiles, with a metallicity well above any seen in the metal-poor sample.

\providecommand{\newl}{\\[0.5mm]}
\begin{deluxetable*}{lllllllllll}
\tablecolumns{17}
\tablecaption{Lyman Limit Systems\label{llstab}}
\tablehead{
\colhead{QSO}&\colhead{$r$\tablenotemark{a}}&\colhead{\zqso}&\colhead{\zlls}&\colhead{$\Delta v$\tablenotemark{b}}&\colhead{$\log \NHI$}&\colhead{$\log U$}&\colhead{\nh}&\colhead{$\ell$}&\colhead{\mh\tablenotemark{d}}\\
&&&&\colhead{(\!\kms)}&\colhead{($\log [\!\cm{-2}]$)}&&\colhead{($10^{-3}$\!\cm{-3})}&\colhead{(kpc)}& }
\startdata
\cutinhead{Metal-Poor Sample}
J080853$-$070940&19.3&3.841&3.545&+80,-80&17.8--18.6&$>-2.57$&$<3.7$&$>28$&$<-2.40^2$\newl
J083832+200142&18.2&3.876&3.476&+220,-130&17.9--18.6&$-2.55^{+0.04}_{-0.05}$&2.8--4.7&30--40&$-1.70^{+0.09}_{-0.07}$\newl
J085944+212511&18.8&3.699&3.263&+190,-110&18.8--19.6&$-2.38\pm0.04$&1.9--3.1&90--110&$-1.99^{+0.06}_{-0.07}$\newl
J110657+081643&19.1&4.268&4.105&\nodata&17.3--17.6&\nodata&$<10.9$&$>1.4$&$<-2.19^1$\newl
J115321+101112&19.1&4.127&4.038&+70,-70&17.7--19.0&$>-2.22$&$<1.8$&130--190&$<-2.90^2$\newl
J123525+014945&19.1&4.031&3.891&+150,-150&17.8--19.0&$-2.11\pm0.06$&1.0--1.8&260--330&$-2.37^{+0.11}_{-0.10}$\newl
J124957$-$015928&17.6&3.638&3.524&\nodata&17.8--19.0&\nodata&$<10.1$&$>3.4$&$<-2.70^{1,}\tablenotemark{d}$\newl
J125949+162005&19.0&3.707&3.547&+110,-110&17.8--18.6&$-1.94^{+0.07}_{-0.09}$&0.6--1.2&650--890&$-2.92^{+0.15}_{-0.13}$\newl
J130452+023924&18.4&3.651&3.336&+110,-110&17.9--18.7&$-2.08\pm0.09$&0.8--1.7&360--460&$-2.81^{+0.15}_{-0.17}$&\newl
J130452+023924&\phn{}&\phn{}&3.324&+90,-90{}&17.9--18.6&$>-2.29$&$<2.0$&130--160&$<-3.08^2$\newl
J131056+105530&19.0&4.461&4.200&+100,+130&18.2--19.0&$-2.45\pm0.05$&2.6--4.1&50--60&$-2.39^{+0.09}_{-0.10}$\newl
J134723+002158&19.3&4.308&4.229&+100,-100&17.5--18.1&$-2.24^{+0.04}_{-0.05}$&1.6--2.5&80--140&$-2.05^{+0.10}_{-0.08}$\newl
J144027+173038&19.5&3.674&3.566&\nodata&19.2--19.6&\nodata&$<10.0$&$>5.2$&$<-1.68^1$\newl
J144405+165621&18.9&3.745&3.551&+150,-130&17.3--17.6&$-2.05\pm0.12$&1.1--1.7&170--290&$-2.41^{+0.20}_{-0.23}$\newl
J144405+165621&\phn{}&\phn{}&3.471&\nodata&17.5--17.8&\nodata&$<10.1$&$>2.2$&$<-2.21^1$\newl
J155255+145432&19.9&4.105&3.954&\nodata&17.8--18.4&\nodata&$<10.5$&$>3.3$&$<-2.03^1$\newl
J160320+072104&19.5&4.393&4.375&+60,-60&17.8--18.4&\nodata&$<11.8$&$>2.8$&$<-1.92^1$\newl

\cutinhead{Metal-Blind Sample}
J001022--003701&18.4&3.153&3.116&-120,+120&17.8--18.8&$-2.11^{+0.04}_{-0.05}$&1.0--1.7&270--400&$-2.19^{+0.07}_{-0.16}$\newl
J014850--090712&18.0&3.322&2.996&-110,+170&17.8--18.7&$-2.46\pm0.03$&2.4--3.8&40--70&$-2.06\pm0.05$\newl
J030341--002321&17.7&3.229&2.941&-150,+150&18.7--19.2&$-2.14\pm0.02$&1.2--1.8&300--320&$-2.07\pm0.03$\newl
H0449--1325&&3.107&2.997&-70,+70&17.8--18.2&$>-2.55$&$<3.8$&$>30$&$<-2.69^2$\newl
J093153--000051&18.7&3.209&2.927&-150,+150&18.4--19.3&$-2.31^{+0.03}_{-0.04}$&1.8--2.8&120--140&$-2.20^{+0.06}_{-0.05}$\newl
J102509+045246&19.2&3.243&3.130&-70,-70&17.8--18.7&$-2.32^{+0.13}_{-0.12}$&1.3--3.3&100--140&$-2.75^{+0.13}_{-0.12}$\newl
J161545+060852&18.2&3.062&2.988&-110,+160&17.8--19.5&$-2.71^{+0.06}_{-0.08}$&4.1--7.6&10--20&$-2.02^{+0.14}_{-0.11}$\newl
J223819--092106&18.0&3.278&3.127&-160,+170&17.8--19.0&$>-2.60$&$<4.1$&$>21$&$>-0.75$\newl
J233446--090812&18.0&3.351&3.226&-90,+100&17.8--18.6&$-2.05^{+0.07}_{-0.06}$&0.8--1.5&380--520&$-2.71^{+0.11}_{-0.15}$\newl
UM184&&3.021&2.929&-110,+110&18.5--19.2&$-2.51^{+0.01}_{-0.02}$&2.9--4.3&40--50&$-1.78^{+0.03}_{-0.05}$

\enddata
\tablenotetext{a}{Quasar r-band magnitude}
\tablenotetext{b}{Velocity width of absorbers, based on metal detections. Dots denote that all lines had no absorption.} %
\tablenotetext{c}{Superscripts indicate Type 1 and Type 2 limits}
\tablenotetext{d}{Using column density limits from the MIKE+ FIRE spectra, we find $\mh<-2.90$.}
\end{deluxetable*}

\begin{figure}[t]
\plotone{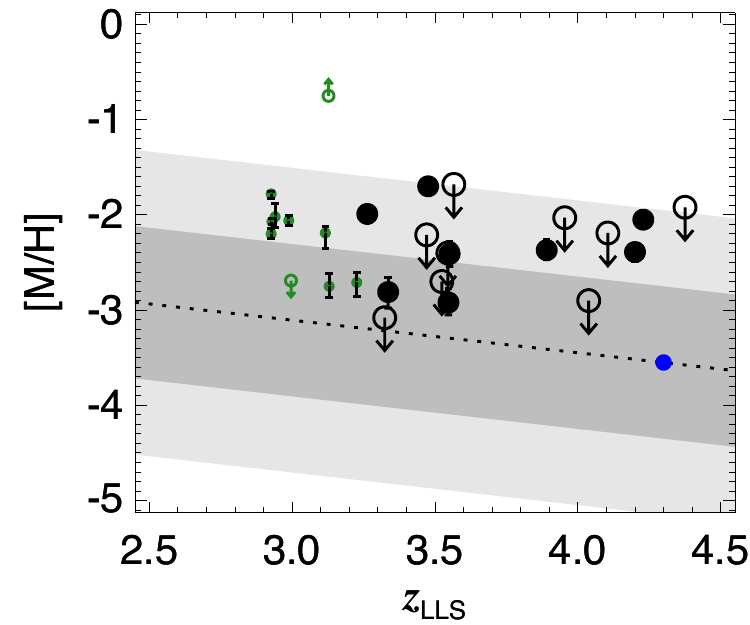}
\caption{Metallicity as a function of redshift for the metal-poor  (black points) and the metal-blind (green) samples. We maintain this color scheme throughout. Arrows indicate upper limits The blue point is the IGM, with shaded regions showing the $1\sigma$ and $2\sigma$ limits. Both samples are consistent with having a fraction of their metallicities drawn from the IGM.}
\label{m_z2}
\end{figure}

In Figures \ref{m_histo} and \ref{m_z2}, we display a histogram of the metallicities and a comparison of the LLS metallicities with the IGM metallicity as a function of redshift. For both the metal-poor and metal-blind samples, measured metallicities range from $\sim\!-2.8$ to $\sim\!-1.8$. Only considering detections, the metal-poor(blind) sample has median metallicity of --2.21(--2.13). Although the distributions for metallicities we were able to measure are not strikingly different, the metal-poor sample has a much larger fraction of systems for which we were only able to assign an upper limit to the metallicity.

\cite{2011ApJ...738..159S} find at $z$=2.4(4.3), the IGM has an abundance of \xh{C}=--2.90(--3.55)$\pm0.8$ dex, as indicated by the shaded region in Figure \ref{m_z2}. All of the LLSs in both our samples have metallicities within $2\sigma$ of the diffuse IGM metallicity, and several of the systems are in very good agreement, having $\mh\!\sim-3.0$ at $z\sim3.5$. Considering the large fraction of the metal-poor sample constituting metallicity upper limits, this suggests that the gas comprising a significant fraction of these absorbers has not cycled through a galaxy; if the corresponding absorbers represent circumgalactic material, they would likely be accreting onto the galactic disk rather than being expelled.

We also note that the detections and upper limits are not well stratified---although one might expect 
upper limits to fall below the detections, many of the measured metallicities are below the upper limits.
This results from detections constraining the gas as more highly ionized
(which typically corresponds to a lower metallicity for the ions we
consider) and could mean systems with upper limits likely have
even smaller metallicities. Additionally, measuring Type I upper limits at the smallest viable value of \NHI\ and $\log U$ results in conservative limits. If we instead use either the median value of allowed \NHI\ or the median $\log U$ from systems with detections, Type 1 limits become stricter by up to $\sim0.5\!$ dex.

\subsection{Ionization Parameters}

\begin{figure}[t]
\plotone{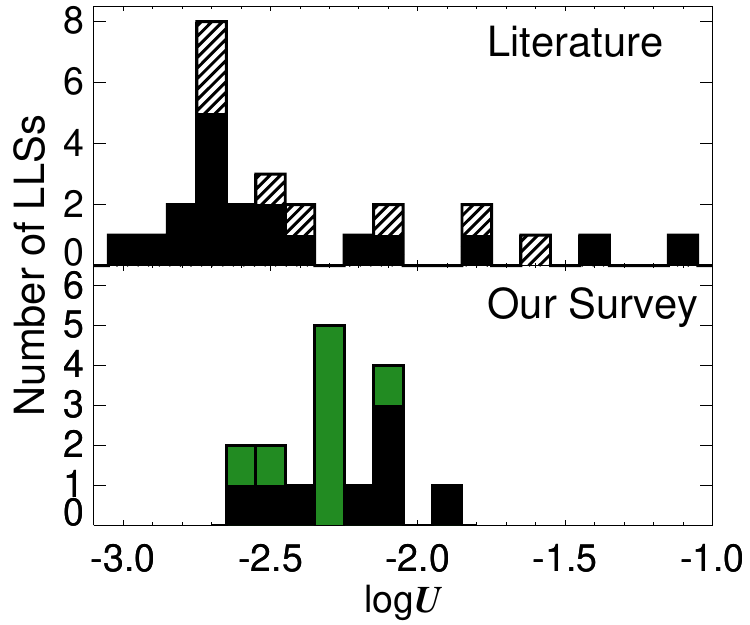}
\caption{Ionization parameters derived for both of our samples and from the literature. In our work, black and green correspond to the metal-blind and metal-poor samples, respectively. LLSs giving metallicity limits are not included. In the literature histogram, the black corresponds to \citet[][and references therein]{2011Sci...334.1245F}, and the hashed region is absorbers from \citet{2015MNRAS.446...18C}, who perform an analysis similar to ours. The literature sample is generally at lower redshift,and contains numerous systems with \NHI\ lower than our sample.}
\label{u_histo}
\end{figure}

Larger values of the ionization parameter $U$ generally correspond to lower derived abundances for the ions we considered, so systematically overestimated ionization parameters would drive our results to artificially low metallicities. \citet{2011Sci...334.1245F} compiled
results from published high-redshift LLSs ($z\!>$1.5, their Figure S6)
and determine that all systems have $-3<\log U<-1$, with most systems in
the range of $-3<\log U<-2.5$. These systems generally have smaller
redshifts than our sample. For all of the absorbers with detected heavy-element absorption in our
metal-poor sample, we find $-2.7<\log U<-1.9$,
with most clustered around $\log U=-2.3$ (Figure \ref{u_histo}).

There are several plausible explanations for the moderately larger
ionization parameters we measured. The ionization parameters in
\citet{2011Sci...334.1245F} were compiled from studies using varying
modeling techniques and ionizing radiation spectra. In addition, the complete literature sample comprises a
relatively small sample, roughly the same size as our survey, often with redshifts and \ion{H}{1} column densities different than our sample, suggesting that they may not form a uniformly selected comparison group for our
results. Indeed, \citet{2011Sci...334.1245F} used the values from the
literature only to show that $\log U=-3$ is a justifiable lower
limit. The absorbers in \citet{2015MNRAS.446...18C} are at $z=2.5$, and have sub-LLSs \ion{H}{1} column densities.

 \citet{2013ApJ...770..138L} studied absorbers with  $16.2<\log \NHI<18.5$ at $z\lesssim{1}$ and found $\log{U}=-3.3\pm0.6$.
 \citet{2014ApJ...792....8W} looked at LLSs located near $L^{\ast}$ galaxies at $z\lesssim0.2$ \citep[overlapping the
sample from][]{2013ApJ...770..138L} and measured a mean ionization
parameter $\log U=-2.8$. A trend of LLS ionization parameter moderately increasing with redshift is supported by the studies previously mentioned and our sample, although the total sample size is small.

The hydrogen density, \nh, for each LLS follows from the ionization parameter $U$ from Equation \ref{eq:u}.
To calculate $\Phi$, we used Equation \ref{eq:phi} using the shape of the UVB spectrum from \cite{{2012ApJ...746..125H}},
renormalized to match the observed \ion{H}{1} photoionization rates from \cite{2013MNRAS.436.1023B}. Under the uniformity assumption, the sizes of the gas clouds can be estimated via $\ell=\NHI/(\chi_\mathrm{H\,I}\nh)$, where the neutral hydrogen ionization fraction $\chi_\mathrm{H\,I}$ is output by the ionization models. In Table \ref{llstab} we list the range of densities corresponding to the ionization parameter posterior distribution, and the range of sizes that follow assuming the central value of the $\log U$ over the range of viable \NHI. Typical densities are of order $10^{-3}\cm{-3}$, and typical lengths are a few tens to hundreds of kiloparsecs, with a median of 160 kpc.

\begin{figure}[t]
\plotone{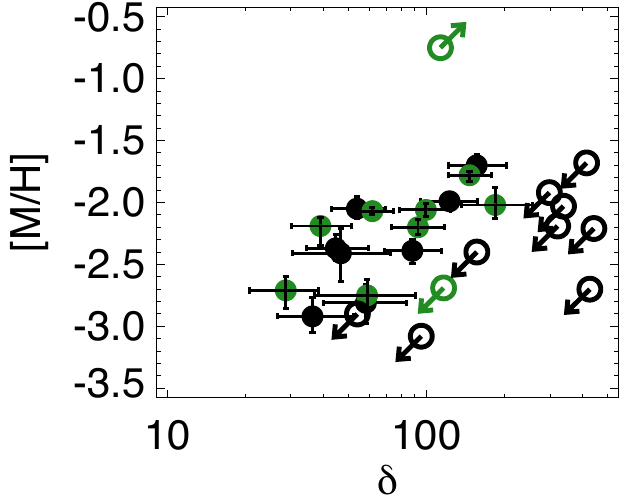}
\caption{Metallicities and overdensities derived from ionization modeling. Uncertainties on the ionization parameter (which corresponds to \nh) and \ion{H}{1} photoionization rates from \cite{2013MNRAS.436.1023B} used to renormalize the ionizing spectrum contribute roughly equally to the overdensity uncertainties. The large limiting overdensities on the Type 1 upper limits suggest the actual metallicities may be much lower.}
\label{odens}
\end{figure}

In Figure \ref{odens}, we plot \mh\ over the overdensities $\delta=\nh/\onh$ of the
absorbers, where \onh\ is the
cosmic mean baryon density at the redshift of the absorber:
$\onh=\Omega_b\rho_c(1+z)^3/m_p$, where $\Omega_b$=0.04 is the cosmic baryon density relative to 
$\rho_c$, the critical density of the universe, and $m_p$ is the proton mass.
The systems with detections have typical overdensities ranging from
$\delta\sim10$--100. This figure also portrays how conservative the
limits are: the Type 1 upper limits correspond to limiting overdensities of at least
two-fold greater than the detections, implying the
metallicities may be significantly lower than the limits we adopted.

\subsection{Effect of High-Resolution and Infrared Spectrum}
\begin{figure*}[t]
\plotone{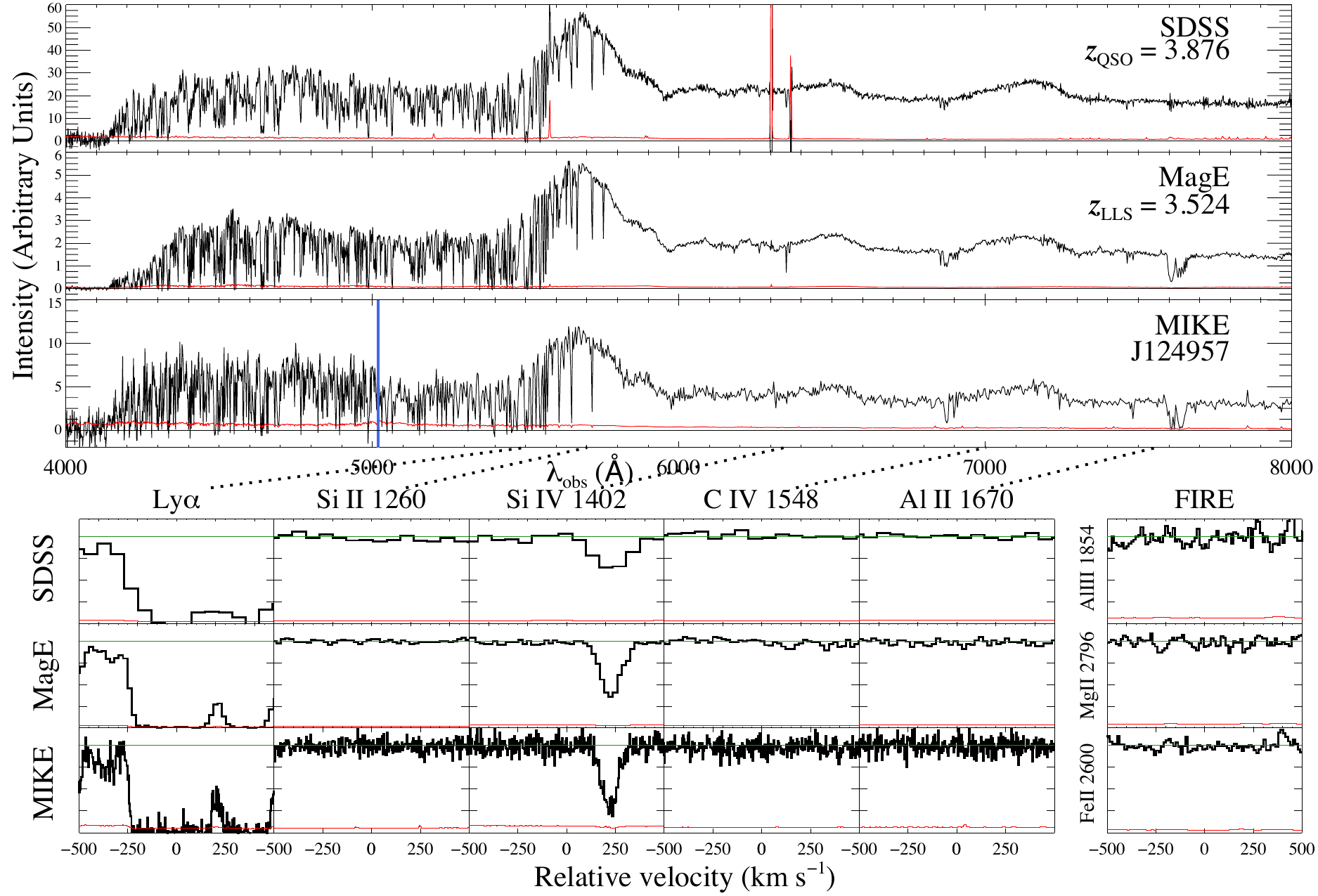} 
\caption{{\bf Top:} Comparison of SDSS, MagE, and MIKE data over a portion of the
  J124957 spectrum. The MIKE spectrum has a blue line indicating
  the transition from the blue and red arms of the instrument. {\bf Bottom:}
  Portions of the normalized spectra around expected LLS
  lines, showing no absorption. The absorption offset by 250\kms\
  in the \ion{Si}{4} panel is an interloping \ion{C}{4} line at
  $z=3.102$. To the right are three cutouts of the FIRE spectrum, also showing no absorption.}
\label{compare_full}
\end{figure*}

The high-resolution MIKE spectrum of J124957 confirmed the lack of
metal lines in the MagE spectrum of the same object was not a
resolution effect, and the IR FIRE spectrum likewise shows no absorption at expected locations.
In Figure \ref{compare_full} we compare the SDSS, MagE,
and MIKE spectra for J124957, as well as showing cutouts of several portions of the FIRE spectrum where absorption is expected. It is evident that the four-fold increase in resolution provided by MIKE does not
reveal any weak lines in this particular case. 

We obtained stricter column density upper limits with the high-resolution spectrum,
allowing us to measure a metallicity upper limit of $\mh<-2.90$,
0.2\,dex less than the upper limit measured with MagE. The limits we measure from the FIRE spectrum are not as strict, since it has lower resolution. When we use column density limits from both MIKE and FIRE, the ions measured with FIRE do not influence the
result. Using only ions from the FIRE spectrum, we find an upper limit of $\mh<-2.60$.

It remains to be seen if high-resolution observations could reveal weaker lines in
other examples, but in light of current sensitivities and to maintain
consistency among the sample, we use the metallicity limit obtained
from the MagE spectrum in subsequent analysis.

\subsection{Survival Analysis}\label{surv}

A complete description of the distribution of metallicities needs to
incorporate information provided by both detections and upper
limits. To that end, we employ survival analysis methods developed to
deal with censored data sets containing a mixture of measurements and
limits.

For univariate data, the Kaplan-Meier estimator provides a
general, non-parametric maximum-likelihood estimate of the population from which
a censored sample was drawn. For details on the application of
the Kaplan-Meier method to similar datasets, see
\citet[][]{2004ApJ...606...92S} and references therein.

Briefly, the Kaplan-Meier method constructs a cumulative distribution
function for the sample, handling ambiguities introduced by upper
limits by only including them in probability calculations when they
are guaranteed to be unambiguous. For example, an upper limit of
$\mh<-2.5$ is guaranteed to be less than $\mh=-2.4$, but may not be less
than $\mh=-2.6$ and will not be treated as such. The resulting
cumulative distribution is a piece-wise function that remains constant
at upper limits and jumps at detections.

The Kaplan-Meier method requires the sample to satisfy two
criteria. First, the upper limits must be independent. This is
clearly true for our sample, where each measurement is drawn from a
different absorber and most are from different sightlines. Second,
the probability of a measurement being censored must be uncorrelated
with the value of the measurement itself. While our sample likely does
not strictly meet this criterion, since lower metallicity systems are
more likely to result in non-detections, there is a characteristic to
our survey that preserves the randomness of the censoring: all
targets were selected using the same criterion, so the priors on metallicity are uniform across the sample.
Hence, the selection method should
adequately randomize the censoring. Also, since an LLS  observation resulting in a metallicity limit partially depends on the signal-to-noise ratio of the spectrum, the brightness of the quasar and observing conditions also serve as randomizing factors. From the discussion in Section \ref{sec:nhidep}, metal column densities generally do not depend strongly on \NHI, so \NHI\ should not bias whether or not a system gives a metallicity upper limit. Since the metallicity measurements and limits we find are not segregated such that all limits fall below detections, the Kaplan-Meier method is applicable.

\begin{figure}[t]
\plotone{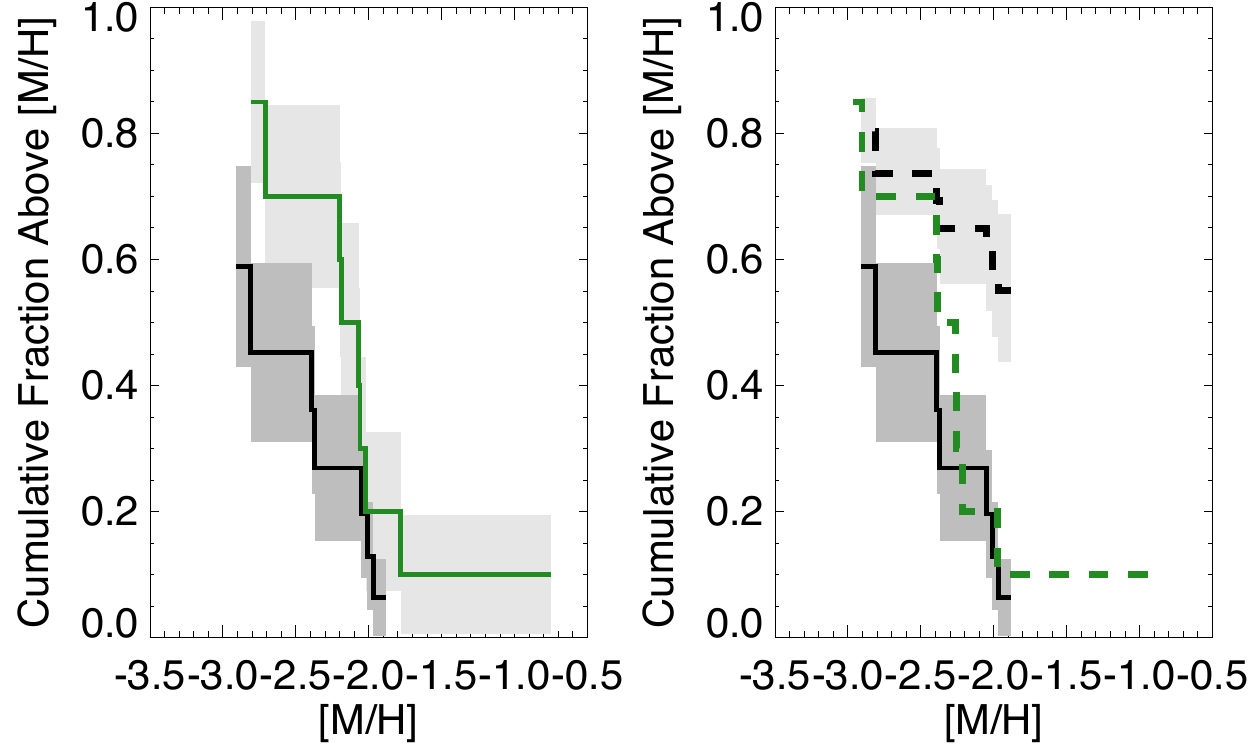} 
\caption{ {\bf Left:} Cumulative distribution functions for the metal-poor (black) and metal-blind (green) LLS samples, constructed using the Kaplan-Meier Estimator. {\bf Right:} Estimates of the full LLS CDF at $z\sim3.7$ extrapolated from the metal-poor (black-dashed) and metal-blind (green-dashed) samples. They disagree at high metallicity since the metal-poor sample does not probe this region.}
\label{asurv}
\end{figure}

We calculate the Kaplan-Meier distributions for our samples using ASURV
Rev 1.2 \citep{1990BAAS...22..917I,
1992BAAS...24..839L}, which implements the methods discussed
in \cite{1985ApJ...293..192F}. The resulting cumulative distributions
are shown in Figure \ref{asurv}.

We also extrapolate our results for each sample independently to estimate the
entire LLS population of $z\sim3.73$, the mean redshift of the metal-poor sample. Since
the metal-blind sample is at lower redshift and metallicities
evolve with redshift, we can apply a shift to the entire cumulative distribution function (CDF) to bring
it to the same redshift as the metal-poor sample. Taking the mean
slope of the IGM and DLA metallicity with redshift, $\mh\propto0.28z$ (see Section \ref{sec:distr}), we shift the metal-blind CDF by the difference between the mean redshifts of the
samples.

Alternatively, we may estimate the full LLS CDF from
the metal-poor sample {\em if} we assume all systems with metal
lines detected in SDSS have higher metallicity than those that do not.
This assertion is demonstrably false for some individual cases, but
lacking the (forthcoming) fully unbiased set of \ion{H}{1}-selected LLS metallicities, it can
represent a first attempt at generalization. In this case we may
construct the CDF for the entire SDSS sample as:
\begin{align}
P_{\rm{SDSS}}&=xP_{\rm{MagE}}+(1-x)\nonumber\\
x&=\frac{\mbox{\footnotesize Number of SDSS LLSs meeting metal-blind criteria}}{\mbox{\small Number of SDSS LLSs}}\nonumber
\end{align}

In other words, we scale the CDF by the fraction of LLSs matching our
selection criterion ($x$), then add to it the fraction of LLSs 
that do not. Since, under our assumption, all LLSs not meeting our
criteria have metallicities larger than those that do, this approximates the CDF for
metallicities in the range probed by our metal-poor sample. In our scan of the SDSS spectra,
we found $x=0.48$. We stress that variations in signal-to-noise and ionization fractions may result
in some SDSS LLSs that failed to meet our initial selection criteria having
lower metallicity than others that did.

In Figure \ref{asurv} we compare these extrapolations to
the metal-poor CDF and to each other. They diverge at high
metallicity, because the assumptions on the extrapolation from the
metal-poor sample place an unrealistic floor on the corresponding CDF,
which is most relevant at higher metallicity. They are in fairly good
agreement for metallicities below $\mh=-2.5$; both extrapolations
suggest $\sim20\%$ of LLS at $z\sim3.73$ have $\mh<-2.5$, a value
roughly $1\sigma$ above the measured IGM abundance.

\section{Discussion}

In this section, we leverage our dataset to extract physical and
cosmological details concerning low-metallicity LLSs
and compare with expectations for cold flows from
simulations. Since our sample size is small, our intention is to establish
an order-of-magnitude for several key properties.

\begin{deluxetable*}{ccccccccc}
\tablecaption{Metallicity Distribution Parameters}
\tablehead{\colhead{}&\colhead{$z$}&\colhead{$\mh_\mathrm{IGM}$}&\colhead{$\sigma_\mathrm{IGM}$}&\colhead{$\mh_\mathrm{DLA}$}&\colhead{$\sigma_\mathrm{DLA}$}&\colhead{$\figm$}&\colhead{$68\%$ c.i.}&\colhead{$95\%$ c.i.}}
\startdata
Metal-poor&3.73&-3.36&0.8&-1.69&0.48&0.71&0.60--0.84&0.44--0.96\\
Metal-blind&3.04&-3.12&0.8&-1.49&0.52&0.48&0.34--0.62&0.18--0.79
\enddata
\label{figm_table}
\end{deluxetable*}

\subsection{Interpreting the Distribution in the Context of Cold--Flows}\label{sec:distr}

In order to assess whether or not low-metallicity LLSs are consistent
with being cold flows, we compare the metallicity CDFs of both the
metal-poor and metal-blind samples to a toy model parent CDF. Motivated by the bimodal
metallicity distribution found at low redshift
\citep{2013ApJ...770..138L}, we assume a mixed Gaussian model
 where the absorbers are drawn from a combination of two different
 parent populations, one being the IGM
 (representing potential accretion flows) and the other having more
 highly enriched gas that has been polluted by a local host galaxy.
 We refer to absorbers drawn from the IGM distribution as cold-flow candidates (CFCs).
 
We assume the parent distribution in \mh\ for CFCs is the same as the IGM's, which we interpolate from the measurements of \cite{2011ApJ...738..159S} to the mean redshift of the sample. Note that this study used the same ionizing background spectrum to measure metallicities, so any systematics from uncertainty in a particular realization of the UVB spectrum are common to both studies.

The parent \mh\ distribution of enriched CGM gas is less clear, but for this study we associate this phase with DLA abundances. The exact physical structures giving rise to DLAs are neither fully understood nor expected to be uniform, but they are thought to be locally enriched. Since DLAs have systematically lower abundances than \ion{H}{2} regions measured in emission \citep[e.g.,][]{2015ApJ...799..138S}, this is a conservative choice to represent the non-CFC branch (using larger metallicities for this branch would give a larger fraction of CFCs). We model enriched gas using a lognormal distribution with parameters given by DLA
measurements, since DLAs are thought to originate from gas in galaxies \citep[e.g.,][]{2011ApJ...736...48R}.
We use DLA metallicities from \citet[][and references therein]{2012ApJ...755...89R}. Compared to LLSs, DLA metallicities have been analyzed
more extensively since DLAs are predominantly neutral and tend to have small ionization corrections that do not require modeling. In addition, the Ly$\alpha$ damping wings allow for
accurate \ion{H}{1} column densities in moderate-resolution
spectra. 

To compare with the metal-poor (blind) sample we average over all DLAs between $z=
3.26$--4.37 (2.90--3.25). The model metallicity probability
distribution is then
\[p(\mh)=\figm\,p_\mathrm{IGM}(\mh)+(1-\figm)p_\mathrm{DLA}(\mh)\]
where $p_\mathrm{IGM}$ and $p_\mathrm{DLA}$ are Gaussian metallicity distributions with parameters summarized in Table \ref{figm_table}, and \figm\ is the fraction of model distribution drawn
from the IGM that we are estimating (i.e., the fraction of CFCs).

\begin{figure}[t]
\epsscale{0.75}
\plotone{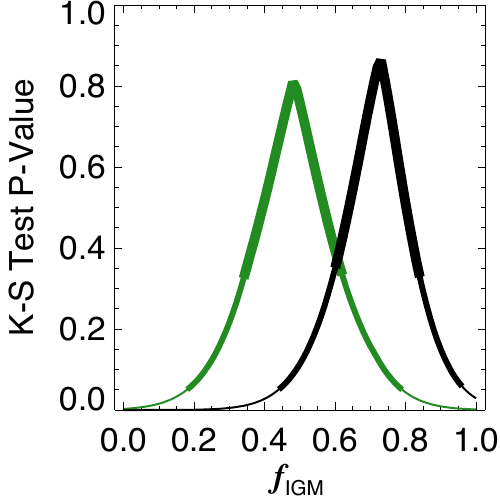} 
\caption{Kolmogorov-Smirnov test P-values as a function of \figm for the metal-poor (black) and metal-blind (green) samples. The null hypothesis is that the observed CDF is drawn from a parent sample having a fraction \figm\ of its metallicities coming from the IGM distribution. Thick bars denote the 68 and 95\% confidence intervals}
\label{ks}
\end{figure}

We performed Kolmogorov-Smirnov tests to find which IGM fractions are allowed by the
 two measured distributions (Figure \ref{ks}). We list the values of \figm\ allowed within $68$ and $95\%$ confidence in Table \ref{figm_table}.

Performing a least-squares fit, we find a best fit to the CDFs with
$\figm=0.71$ and 0.48 for the metal-poor and metal-blind samples, respectively. We tested several different likelihood functions and found approximately the same maximum likelihood values for \figm. We adopt these best-fit 
\figm\ values, with errors given by the $68\%$ confidence intervals
for the remainder of the paper. We caution that the small sample sizes
enable intrinsic variation within the LLS population to appreciably
influence the results.

Since $48\%$ of the SDSS sample meets our metal-poor 
criteria, assuming only systems passing our initial cuts can
be cold-flow candidates implies that the range of acceptable values
for $\figm$ for the entire $z\sim3.7$ LLS population is
$0.34\pm0.06$. This is somewhat less than \figm\ for the
metal-blind sample, suggesting that the assumption regarding the SDSS sample may be
questionable, as discussed in Section \ref{surv}, although sample variance may also account for the disagreement.

\begin{figure}[t]
\plotone{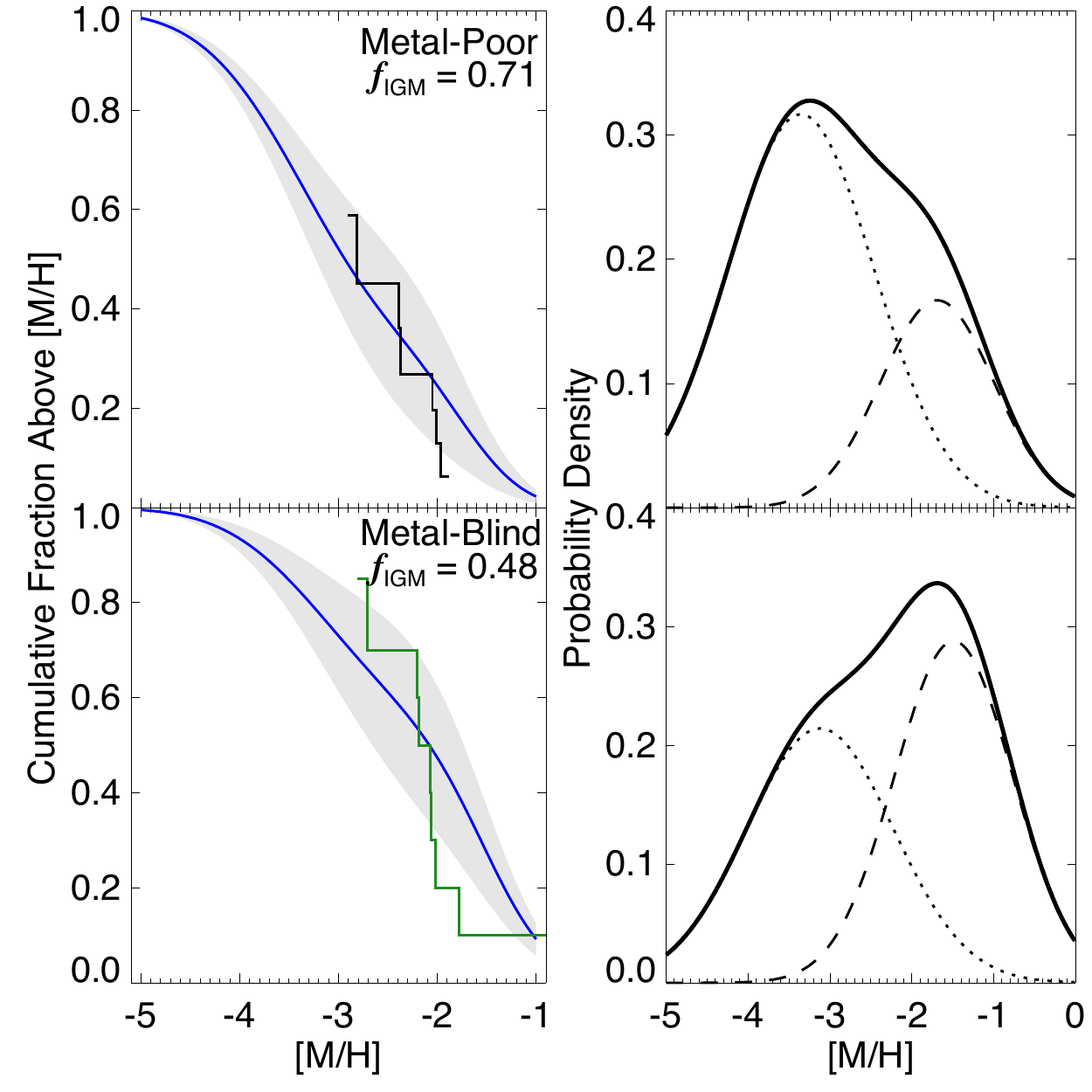} 
\caption{{\bf Left:} Model CDFs corresponding to the best-fit \figm\ (blue) overlaid onto measured CDFs. The shaded region encompasses $68\%$ confidence on \figm. {\bf Right:} PDF corresponding to the best-fit \figm\ and the contributions from the IGM (dotted) and DLA (dashed) distributions.}
\label{figm}
\end{figure}

In Figure \ref{figm}, we show the best-fit CDF and $68\%$
confidence intervals for both samples, as well as the best-fit probability distribution function with the
relative contributions from the enriched and unenriched parent
populations. 

\subsection{Mass Fraction of Candidate Cold Flows}

Of particular interest is $\Omega_{\rm CFC}$: the mass fraction of the Universe (relative to the critical density) at
$z\sim4$ contained in LLSs with IGM metallicities (i.e., CFCs). This quantity can be compared to simulations to test
whether the global mass contained in cold flows agrees.
Using our measurements and ionization
models we make an order-of-magnitude estimate of this
quantity for comparison with simulations.  

The ratio of the CFC mass density to the cosmological
critical density, $\rho_{\rm c}$, is given by:
\[\Omega_\mathrm{CFC}=\frac{H_0}{c}\frac{\mu m_{\rm H}}{\rho_{\rm c}}\int_0^\infty N_\mathrm{H}(\NHI) f_\mathrm{CFC}(\NHI)d\NHI\]
where $\NH(\NHI)$ is the total hydrogen column density of an absorber, $f_\mathrm{CFC}(\NHI)$ is the frequency distribution of CFCs as a function of neutral hydrogen column density, $\mu$ is the reduced mass of the gas and $m_{\rm H}$ is the mass of the hydrogen atom.

To compute the integral, we need to assume a form for
$f_\mathrm{CFC}(\NHI)$  \citetalias{2010ApJ...718..392P} find
$f(\NHI)$ for $z=3.7$ LLSs can be fit by:
\begin{align}
	f_\mathrm{LLS}(\NHI)=\left\{
		\begin{array}{lr}
			10^{-4.85}\NHI^{-0.8} & 17.5<\log\NHI\le19\\[1em]
			10^{2.75}\NHI^{-1.2} &  19<\log\NHI<20.3
		\end{array}
	\right. \nonumber
\end{align}
To estimate $f_\mathrm{CFC}(\NHI)$, we multiply $f_\mathrm{LLS}$ by the fraction of LLSs that are cold-flow candidates, \figm. (Note \figm\ is the fraction of LLS that are CFCs and $f_\mathrm{CFC}$ is the frequency distribution of these LLSs). We take \figm=0.34$\pm$0.06 from Section \ref{sec:distr}.

We assume a reduced mass $\mu=1.3$, appropriate for absorbers with $75\%$ H and $25\%$ He by mass. The total hydrogen column density is readily found from the \ion{H}{1} column density $N_\mathrm{H}=(n_\mathrm{H}/n_\mathrm{HI})\NHI$,
with the ratio of ionized-to-total hydrogen coming from the ionization model. We use the median value derived from ionization solutions in both of our samples: $\langle n_\mathrm{H}/n_\mathrm{HI}\rangle=0.0063$. The median value for each sample separately is similar. Due to our small sample size, we cannot adequately measure and include in the calculation any variation in $f_\mathrm{CFC}$ and $\langle n_\mathrm{H}/n_\mathrm{HI}\rangle$ with \ion{H}{1} column density.

We also need to set bounds on the integral. For the lower bound, we use $\log\NHI=17.5,$ the stated sensitivity limit of the LLS survey in \citetalias{2010ApJ...718..392P}. Although lower column density absorbers are more numerous, larger column density absorbers dominate the mass density, so the result is largely insensitive to the lower bound. As an upper bound we take $\log\NHI=19.5$, a typical value where systems transition from being considered LLSs to sub-DLAs.

With these parameters, we obtain $\Omega_\mathrm{CFC}=0.0017$. Comparing this to the cosmic baryon density, $\Omega_\mathrm{b}=0.04$, we find roughly $5\%$ of baryons at $z\sim3.7$ are contained in cold-flow candidate LLSs. Note this calculation is sensitive to both the maximum column density used in integration and the \ion{H}{1} ionization fraction; increasing (decreasing) the maximum \NHI\ by 0.2 dex increases (decreases) the result by a factor of $\sim\!1.5$, and the \ion{H}{1} ionization fractions for individual systems can vary from the median by a factor of $\sim\!5$. As we discuss in Section \ref{sec:compsim}, our result is fairly consistent with simulated results.

\subsection{Comparison with Other Observations}\label{sec:compobs}

\begin{figure}
\includegraphics[width=3.45in]{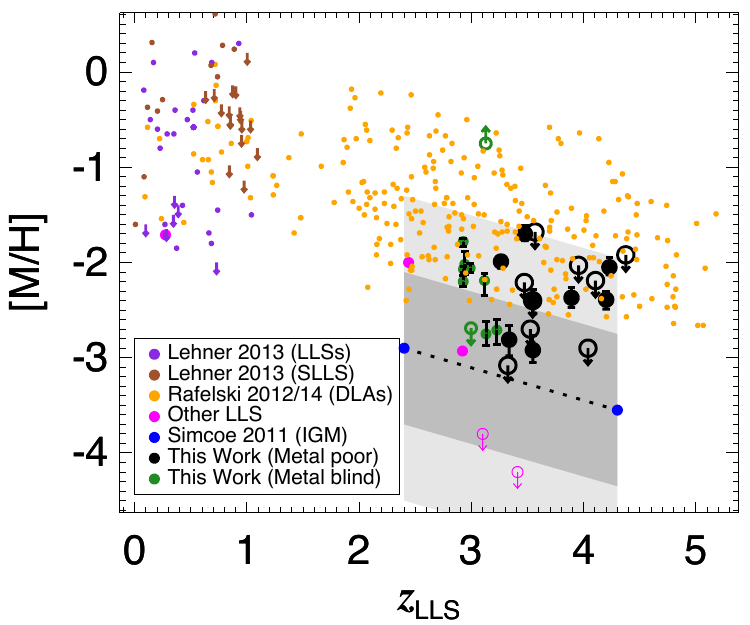}
\caption{Comparison of our results with DLAs (orange) and LLSs (various) from the literature. At high redshift, the DLA and LLS metallicity distributions are more clearly different than at low redshift. The bimodality seen in low-redshift LLS metallicities is not evident at high redshift, although the population of high-redshift LLSs has not been fully explored.}
\label{m_z3}
\end{figure}

In Figure \ref{m_z3}, we compare the metallicities of our samples to DLA
metallicities from \citet[][and references therein]{2012ApJ...755...89R}.
DLAs generally have higher metallicities than both our
metal-poor and metal-blind samples, and there are suggestions that
DLAs have a metallicity floor, which many LLSs are below. It is clear
from this comparison that LLSs and DLAs at high
redshift differ significantly in their metallicity distributions.

Also shown are low-redshift LLSs from
\citet[][and references therein]{2013ApJ...770..138L}. They categorized
their \ion{H}{1} systems as LLS ($16.2\le\log\NHI<19$) or super 
LLS (SLLS, $19\le\log\NHI<20.3$). Their LLS sample is mostly composed of
systems with $\log\NHI<17.5$, the cutoff for our survey, so some
differences are expected.

The low-redshift LLS population clearly exhibits a
metallicity bimodality, with most of the lower-metallicity branch below
most DLAs, although the difference between the LLS
and DLA populations is not as emphasized as at higher
redshift.  While a bimodal model fits our high-redshift
sample well, the two populations blend together more smoothly than
they do at lower redshift, where \citeauthor{2013ApJ...770..138L} see very few systems at
intermediate abundances. This may be a result of many of the lower abundances at high-redshift producing
upper limits rather than measurements.

\nocite{2013ApJ...770..138L, 2012ApJ...755...89R, 2011ApJ...738..159S, 2013ApJ...776L..18C, 2003A&A...397..851L, 2013ApJ...775...78F}

\begin{figure*}[t]
\begin{centering}
\includegraphics[width=3.45in]{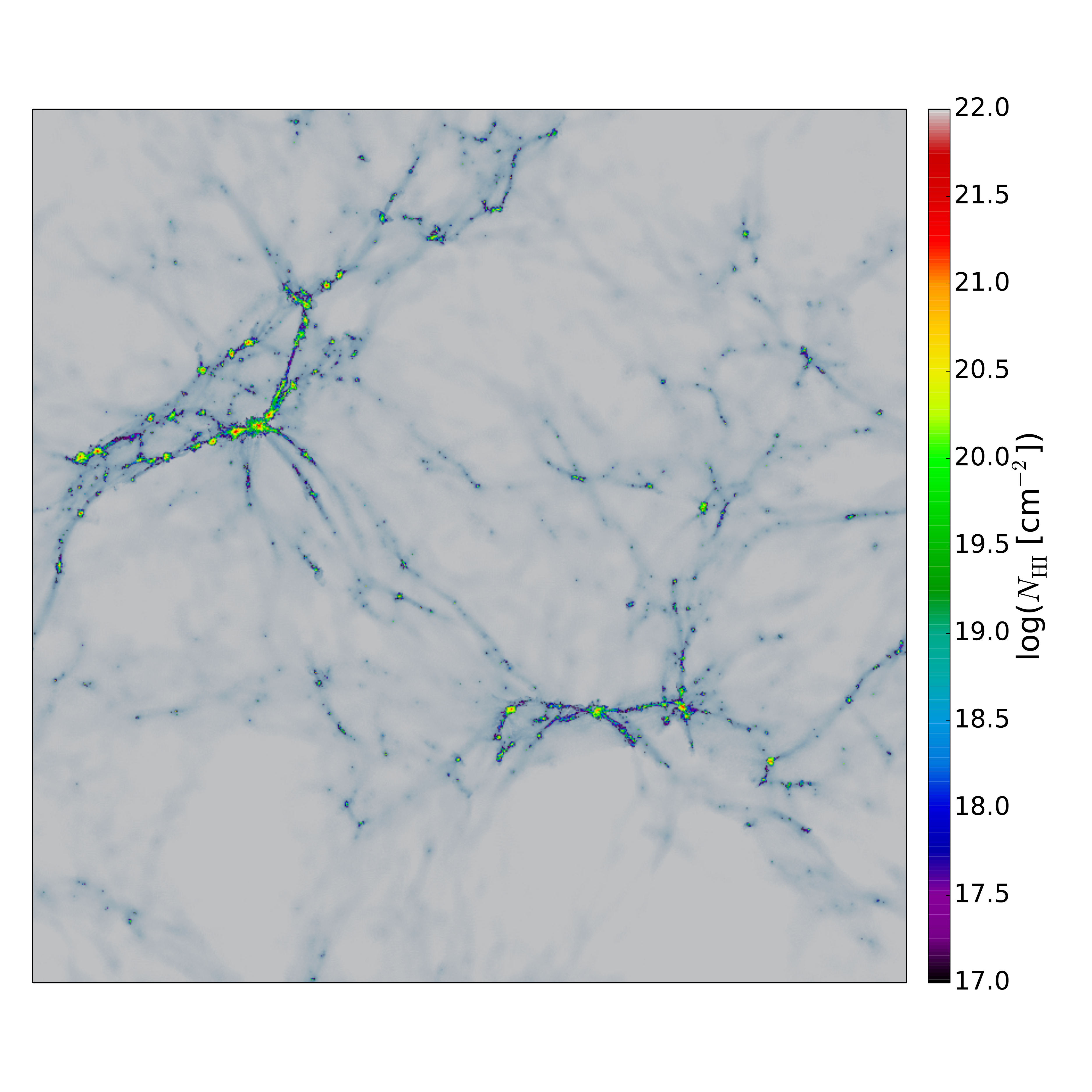}
\includegraphics[width=3.45in]{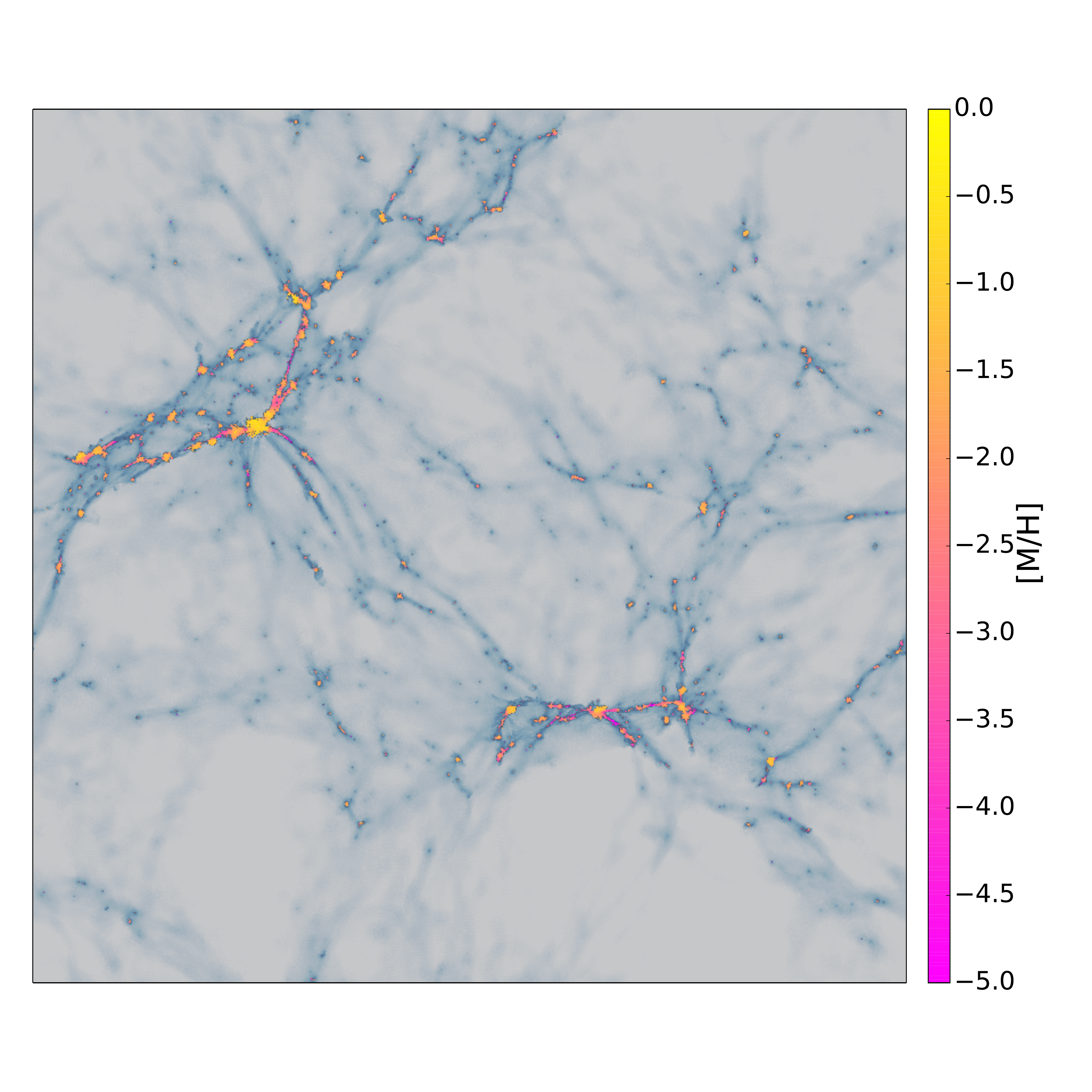}
\caption{\ion{H}{1} column density (left) and metallicity (right) for sightlines with $\log\NHI>17.5$ in a 1 Mpc thick slice of our cosmological simulation. Lower column density material appears as semi-transparent blue. The simulated LLS gas within halos tends to be enriched, with metal-poor LLSs tracing the cosmic web. This suggests either inflowing material is being artificially contaminated by feedback or observed metal-poor LLSs are outside of large galactic halos.}
\label{sims}
\end{centering}
\end{figure*}

Several LLSs drawn from the literature (Table \ref{lit_table}) that
exhibit low metallicities and are claimed as potential evidence of
cold flows are also included in Figure \ref{m_z3}. Our work corroborates the
finding of low-metallicity LLSs and provides some statistical context
for the population from which they are drawn. The two high-redshift
metallicity upper limits are from \cite{2013ApJ...775...78F}. Using
high-resolution spectra, they were able to model and subtract the Ly$\alpha$
forest to obtain column-density upper limits for \ion{C}{3} $\lambda977$ and
\ion{Si}{3} $\lambda1206$, which provide tighter constraints than the ions available
in our medium-resolution spectra (see their Figure S5).

\subsection{Comparison with Simulations}\label{sec:compsim}

By comparing our sample with structures having analogous properties in simulations, we explore the agreement between simulations and observations and gain insight into the nature of metal-poor LLSs. \cite{2011MNRAS.418.1796F} simulated absorption profiles produced by cold flows at $z\sim2.3$ and found that much of the gas is ionized by the UV background, appearing mostly as LLSs. They determined that DLAs have higher metallicities than LLSs and SLLSs, with DLA metallicities fairly consistent with observed systems, and the authors suggested metal-poor LLSs may therefore be an observational signature of cold-flow accretion. Their simulations predict that most of the cold-flow observational signatures are LLSs with $17<\log\NHI<18$, with a peak metallicity for cold stream LLSs of one-hundredth solar.

While our measured metallicities tend to be somewhat lower, enrichment of the IGM between $z=3.7$ and $z=2.3$ may account for some of the difference. An additional difference is that our metal-poor sample is mostly composed of absorbers with $18<\log\NHI<19$, the higher end of the LLS column-density distribution.

A limitation of the simulations employed in \cite{2011MNRAS.418.1796F} and similar studies is the simulated volume.  In \cite{2011MNRAS.418.1796F}, zoom-in simulations of seven halos were considered. This restricts analysis of the covering fraction of LLSs to roughly a few times the virial radius of each galaxy. Although these simulations allowed for a descriptive picture of the neutral-gas content immediately around the seven relatively massive galaxies presented, it remains unclear if a random quasar sightline is more likely to intersect a LLS in the immediate vicinity of one of these systems or elsewhere in the cosmic web. Such questions can only be addressed with larger simulation volumes. Full volume cosmological simulations can complement the \cite{2011MNRAS.418.1796F} analysis by probing the full range of LLSs that are probed through quasar-selected samples.

\begin{deluxetable}{llll}
\tablecaption{Low-Metallicity LLSs in Literature}
\tablehead{\colhead{QSO}&\colhead{\zlls}&\colhead{\mh}&\colhead{Source}} 
\startdata
PG1630+377&0.274&$-1.71\pm0.06$& \cite{2013ApJ...770..138L}\\
J144535+291905&2.44&$-2.0\pm0.17$& \cite{2013ApJ...776L..18C}\\
HE 0940--1050&2.917&$-2.93\pm0.13$& \protect{\cite{2003A&A...397..851L}}\\
J113418+574204&3.411&$<-4.2$& \cite{2013ApJ...775...78F}\\
Q0956+122&3.096&$<-3.8$ &\cite{2013ApJ...775...78F} 
\enddata
\label{lit_table}
\end{deluxetable}

To demonstrate this point, we briefly consider the global distribution of neutral hydrogen in the full-volume
cosmological simulations presented in \citet{2014MNRAS.445.2313B}. These simulations were run using the cosmological hydrodynamical simulation code \texttt{AREPO} \citep{2010MNRAS.401..791S} in a periodic box of size $10\,h^{-1}$\,Mpc.
The simulations contain $512^3$ dark matter particles and a similar number of baryon resolution elements yielding a mass resolution of $1.4\times10^{5}\,M_\odot$---about an order-of-magnitude larger than that presented in \cite{2011MNRAS.418.1796F}. The simulation physics is the same as in in the Illustris Simulation \citep{2014MNRAS.444.1518V} which importantly includes star-formation driven winds at a level that allows for appropriate evolution of the galaxy stellar-mass function \citep{2014MNRAS.438.1985T, 2014MNRAS.445..175G}. Neutral hydrogen fractions are obtained assuming a uniform UV background, with self-shielding corrections \citep{2013MNRAS.430.2427R}.  Complete simulation and post processing details are presented in \citet{2014MNRAS.445.2313B}.

\begin{figure}[b]
\includegraphics{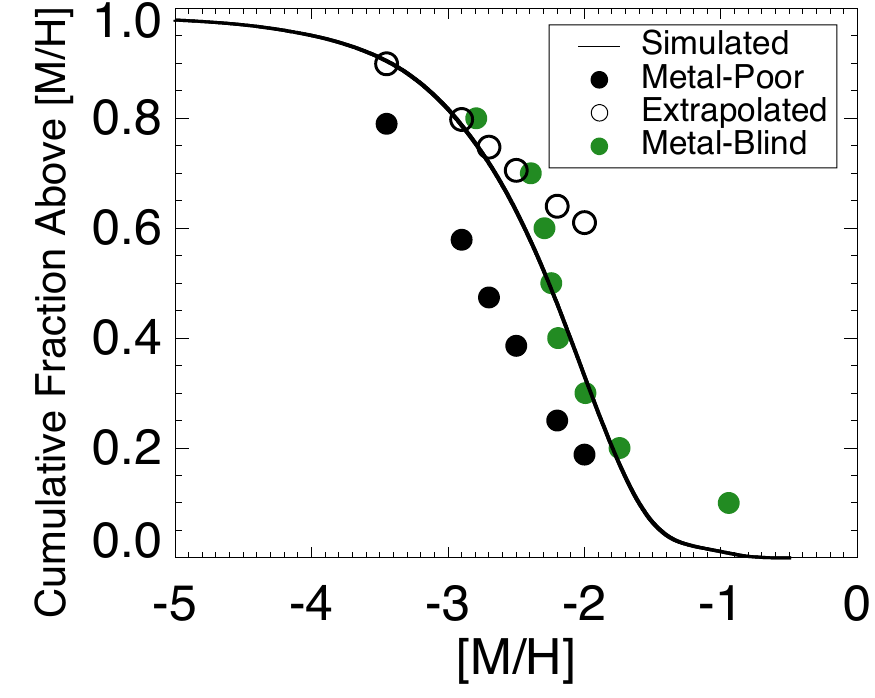}
\caption{Cumulative fraction of LLS abundances from cosmological simulation at $z=3.5$. The metal-blind CDF (shifted to $z=3.5$ according to the redshift evolution in IGM and DLA metallicities) is in fairly good agreement, while the metal-poor CDF is lower, as expected since it is a metallicity biased sample. The full LLS distribution extrapolated from the metal-poor sample as discussed in Section \ref{surv} is in agreement at low metallicities.}
\label{sim_cdf}
\end{figure}

We examine the simulations by applying a similar technique to that used in both \cite{2011MNRAS.418.1796F} and \citet{2014MNRAS.445.2313B}. Specifically, we project the neutral hydrogen and mass-weighted average metallicity onto a two-dimensional grid. We do this only along one projected direction but do not expect our results to be modified if we considered other projections. We employ a grid of 16,000 by 16,000 cells, which results in converged neutral-hydrogen column-density distribution functions. Projecting the full 10\,Mpc box onto a single grid could boost the LLS number count by adding multiple, well-separated low column density systems. To minimize this effect, we use ten slices each with a thickness of 1\,Mpc. We treat each pixel as an independent line of sight.

In Figure \ref{sims}, we show a map of the neutral-hydrogen column density through one slice of the simulated box at $z=3.5$, truncated at $\log\NHI=17.5$ (to emphasize LLSs), and the corresponding map of metallicities. Lower column density material appears as semi-transparent blue in the column-density map. There is
LLS-level column density material tracing the cosmic web extending beyond the virial radii of galaxies in the simulated volume. 

Since we do not have explicit projected offsets for the observed quasar sightlines, it is hard to know which part of the IGM\slash CGM is being probed. It is possible some fraction of our observed CFCs intersect material still in the IGM, well outside of the halo virial radius. In those cases, it is not immediately clear over what timescale the observed neutral gas will fall through the virial radius nor whether it will stay neutral (or be shock heated) as it is accreted. Additionally, most of the simulated LLSs within a halo virial radius are enriched, with the metal-poor LLSs tending to trace the cosmic web. This suggests that either the feedback prescription in the simulation artificially contaminates inflowing material or much of our observed sample is inter-halo material, as opposed to intra-halo.

We can directly address global statistics of the simulated LLS population. Adopting standard column density limits ($17.5 \le \log \NHI < 19.5$), we find roughly $7\%$ of the simulation baryons reside in LLSs, and it has been previously shown that the column density distribution function for neutral hydrogen in these models is reasonably consistent with observations \citep{2014MNRAS.445.2313B}. The simulated LLS CDF at redshift $z=3.5$ is shown in Figure \ref{sim_cdf}.
Our metal-blind sample CDF (when shifted to $z=3.5$)\footnote{The agreement between the unadjusted metal-blind sample and the simulation at $z=3$ is similar.} is in fairly good agreement with the simulated LLS CDF with a peak metallicity for LLSs around one-hundredth solar.  Although we are sampling a significantly larger volume, this result is similar to that presented in \cite{2011MNRAS.418.1796F}. An extrapolation of the full LLS population from the metal-blind sample is in agreement at low metallicities, but diverges at larger metallicities where the simple extrapolation is inadequate.

We consider all LLSs with $\mh<-2.5$ to be CFCs. The fraction of LLSs constituting CFCs is approximately the ratio of the LLS and CFC  mass densities, $\Omega_\mathrm{CFC}/\Omega_\mathrm{LLS}=0.312$. This is in accordance with the observational result for \figm\ extrapolated to the entire LLS population. 

However, we find the derived $\Omega_{\rm{CFC}}$ to be smaller for the simulations by a factor of $\sim2$.
Given that the simulations have both a similar \NHI\ distribution function \citep{2014MNRAS.445.2313B} and metal-poor fraction when compared with observations, it is likely this offset is driven by the applied ionization corrections. In our $\Omega_\mathrm{CFC}$ calculation based on observations, we assumed the hydrogen neutral fraction, $n_\mathrm{H\,I}/\nh$,
 is independent of $\NHI$ due to an insufficient amount of data to measure any such relationship. Since the integrals are heavily weighted towards larger $\NHI$ systems, unaccounted for variations in neutral fraction could systematically and significantly influence the calculation. Hence, the disagreement between simulated and observed values of $\Omega_{\mathrm{CFC}}$ and $\Omega_{\mathrm{LLS}}$ may be resolved by improved observational statistics and does not necessarily undermine the agreement on the fraction of LLSs that are metal poor.

Additionally, we suspect the prevalence of metal-poor LLSs in the simulations can be influenced by (i) the specifics of the adopted feedback model and (ii) the mass and spatial resolution. Further investigations on both of these fronts are warranted, alongside developing an understanding of the dependence of the hydrogen neutral fraction on $\NHI$. Full-volume cosmological simulations that are able to simultaneously reproduce the \NHI\ distribution function as well as the low-metallicity tail of the LLS distribution presented in this paper will be helpful in identifying the fraction of low-metallicity LLSs residing within halo virial radii and, thus, the true mass density of cold flows.

\section{Summary}
We have completed a medium-resolution spectroscopic survey of 17 Lyman-limit systems exhibiting no statistically significant metal-line absorption in the SDSS discovery spectra (${\rm FWHM} \approx 150\kms$) to probe the low abundance end of the LLS population. The main results are as follows:
\begin{enumerate}
\item Five of the LLSs exhibit no statistically
  significant absorption at any of the available metal transitions at MagE
  resolution (${\rm FWHM}=60.7\kms$)
In total we found nine metallicity upper limits, ranging
  from $\mh<-1.68$ to $<-3.08$, with three of the
 upper limits below $\mh=-2.50$. The eight remaining LLSs
  have metallicities ranging from $\mh=-2$ to $-3$ and
  ionization parameters ranging from $\log U=-1.9$ to $-2.6$. The
  median metallicity for the detections is $\mh=-2.21$.
  
  \item A sample of ten LLSs at $z\approx3$ selected blindly with respect to
  metal line absorption exhibits somewhat different properties. Although the median for the systems with measured metallicities is roughly the same at $\mh=-2.13$, only one of the systems has no metal absorption lines. Additionally, this sample contains one LLS with saturated metal lines, leading to a metallicity lower-limit of $\mh>-0.75$.  Taking into the account that over half of the metal-poor sample LLSs have metallicity upper limits, the two samples may have very different metallicity distributions, as demonstrated using survival statistics.
  
  \item LLSs in both samples have typical densities of $(1-5)\times 10^{-3}\cm{-3}$, with corresponding overdensities ranging from 20--200. Length scales span several tens to hundreds of kiloparsecs, with a median of 160 kpc.
    
\item From the cumulative distribution function that results from a
  survival analysis of the detections and limits, the metal-poor sample is consistent with
  $0.71^{+0.13}_{-0.11}$ of the metallicities being drawn from the IGM
  metallicity distribution. Nearly half 
  of the LLSs in SDSS spectra meet our criterion for being metal poor,
  implying that 28--40$\%$ of LLSs at $z=3.2$--4.4 have IGM-consistent metallicities. The metal-blind sample is consistent with an IGM metallicity fraction of $0.48^{+0.14}_{-0.12}$. A
  comparison between LLSs and DLAs shows that have distinct
  metallicity distributions, with many LLSs having metallicities below
  the DLA metallicity floor. 
  
\item We find the cosmic density of low-metallicity LLSs (cold-flow candidates)
  to be $\Omega_\mathrm{CFC}\sim0.0017$, accounting for $\sim5\%$ of the total baryonic mass
  budget at this redshift. This is roughly twice the baryonic fraction of CFCs in simulations, with
  the disagreement likely attributable to limited information of the hydrogen neutral fraction and frequency distribution of low-metallicity LLSs.
 Simulations agree with our observed
  fraction of metal-poor LLSs, although simulations
  call into question what type of gas (inflowing versus IGM) is probed
  along sightlines with metal-poor LLSs. 
  
  \end{enumerate}

This evidence indicates that a statistically significant population of
low-metallicity LLSs exists at redshift $z=3.5$--4.5;
these absorbers have metallicities consistent with being drawn from the IGM
and may therefore be an observational manifestation of filamentary
cold flows predicted by simulations. Observational and archival programs that will increase the moderate-resolution sample
of both the metal-poor and general LLS
populations are underway. This will allow for more in-depth discussion of the metallicity
distribution and cosmological implications, and further coupling to the increasingly more detailed analysis of simulated volumes,
mapping the distribution and flow of unprocessed LLS gas relative to star-forming galaxies.

\bibliography{cflls2015}

\begin{thebibliography}{}
\expandafter\ifx\csname natexlab\endcsname\relax\def\natexlab#1{#1}\fi

\bibitem[{{Becker} \& {Bolton}(2013)}]{2013MNRAS.436.1023B}
{Becker}, G.~D., \& {Bolton}, J.~S. 2013, \mnras, 436, 1023

\bibitem[{{Bernstein} {et~al.}(2003){Bernstein}, {Shectman}, {Gunnels},
  {Mochnacki}, \& {Athey}}]{2003SPIE.4841.1694B}
{Bernstein}, R., {Shectman}, S.~A., {Gunnels}, S.~M., {Mochnacki}, S., \&
  {Athey}, A.~E. 2003, in Society of Photo-Optical Instrumentation Engineers
  (SPIE) Conference Series, Vol. 4841, Instrument Design and Performance for
  Optical/Infrared Ground-based Telescopes, ed. M.~{Iye} \& A.~F.~M.
  {Moorwood}, 1694--1704

\bibitem[{{Bird} {et~al.}(2014){Bird}, {Vogelsberger}, {Haehnelt}, {Sijacki},
  {Genel}, {Torrey}, {Springel}, \& {Hernquist}}]{2014MNRAS.445.2313B}
{Bird}, S., {Vogelsberger}, M., {Haehnelt}, M., {et~al.} 2014, \mnras, 445,
  2313

\bibitem[{{Bochanski} {et~al.}(2009){Bochanski}, {Hennawi}, {Simcoe},
  {Prochaska}, {West}, {Burgasser}, {Burles}, {Bernstein}, {Williams}, \&
  {Murphy}}]{2009PASP..121.1409B}
{Bochanski}, J.~J., {Hennawi}, J.~F., {Simcoe}, R.~A., {et~al.} 2009, \pasp,
  121, 1409

\bibitem[{{Bordoloi} {et~al.}(2014){Bordoloi}, {Tumlinson}, {Werk},
  {Oppenheimer}, {Peeples}, {Prochaska}, {Tripp}, {Katz}, {Dav{\'e}}, {Fox},
  {Thom}, {Ford}, {Weinberg}, {Burchett}, \& {Kollmeier}}]{2014ApJ...796..136B}
{Bordoloi}, R., {Tumlinson}, J., {Werk}, J.~K., {et~al.} 2014, \apj, 796, 136

\bibitem[{{Bournaud} \& {Elmegreen}(2009)}]{2009ApJ...694L.158B}
{Bournaud}, F., \& {Elmegreen}, B.~G. 2009, \apjl, 694, L158

\bibitem[{{Bournaud} {et~al.}(2009){Bournaud}, {Elmegreen}, \&
  {Martig}}]{2009ApJ...707L...1B}
{Bournaud}, F., {Elmegreen}, B.~G., \& {Martig}, M. 2009, \apjl, 707, L1

\bibitem[{{Churchill} {et~al.}(2015){Churchill}, {Vander Vliet},
  {Trujillo-Gomez}, {Kacprzak}, \& {Klypin}}]{2015ApJ...802...10C}
{Churchill}, C.~W., {Vander Vliet}, J.~R., {Trujillo-Gomez}, S., {Kacprzak},
  G.~G., \& {Klypin}, A. 2015, \apj, 802, 10

\bibitem[{{Cooksey} {et~al.}(2013){Cooksey}, {Kao}, {Simcoe}, {O'Meara}, \&
  {Prochaska}}]{2013ApJ...763...37C}
{Cooksey}, K.~L., {Kao}, M.~M., {Simcoe}, R.~A., {O'Meara}, J.~M., \&
  {Prochaska}, J.~X. 2013, \apj, 763, 37

\bibitem[{{Crighton} {et~al.}(2013){Crighton}, {Hennawi}, \&
  {Prochaska}}]{2013ApJ...776L..18C}
{Crighton}, N.~H.~M., {Hennawi}, J.~F., \& {Prochaska}, J.~X. 2013, \apjl, 776,
  L18

\bibitem[{{Crighton} {et~al.}(2015){Crighton}, {Hennawi}, {Simcoe}, {Cooksey},
  {Murphy}, {Fumagalli}, {Prochaska}, \& {Shanks}}]{2015MNRAS.446...18C}
{Crighton}, N.~H.~M., {Hennawi}, J.~F., {Simcoe}, R.~A., {et~al.} 2015, \mnras,
  446, 18

\bibitem[{{Dekel} {et~al.}(2009{\natexlab{a}}){Dekel}, {Sari}, \&
  {Ceverino}}]{2009ApJ...703..785D}
{Dekel}, A., {Sari}, R., \& {Ceverino}, D. 2009{\natexlab{a}}, \apj, 703, 785

\bibitem[{{Dekel} {et~al.}(2009{\natexlab{b}}){Dekel}, {Birnboim}, {Engel},
  {Freundlich}, {Goerdt}, {Mumcuoglu}, {Neistein}, {Pichon}, {Teyssier}, \&
  {Zinger}}]{2009Natur.457..451D}
{Dekel}, A., {Birnboim}, Y., {Engel}, G., {et~al.} 2009{\natexlab{b}}, \nat,
  457, 451

\bibitem[{{Elmegreen} {et~al.}(2007){Elmegreen}, {Elmegreen}, {Ravindranath},
  \& {Coe}}]{2007ApJ...658..763E}
{Elmegreen}, D.~M., {Elmegreen}, B.~G., {Ravindranath}, S., \& {Coe}, D.~A.
  2007, \apj, 658, 763

\bibitem[{{Faucher-Gigu{\`e}re} \& {Kere{\v s}}(2011)}]{2011MNRAS.412L.118F}
{Faucher-Gigu{\`e}re}, C.-A., \& {Kere{\v s}}, D. 2011, \mnras, 412, L118

\bibitem[{{Faucher-Gigu{\`e}re} {et~al.}(2008){Faucher-Gigu{\`e}re}, {Lidz},
  {Hernquist}, \& {Zaldarriaga}}]{2008ApJ...688...85F}
{Faucher-Gigu{\`e}re}, C.-A., {Lidz}, A., {Hernquist}, L., \& {Zaldarriaga}, M.
  2008, \apj, 688, 85

\bibitem[{{Feigelson} \& {Nelson}(1985)}]{1985ApJ...293..192F}
{Feigelson}, E.~D., \& {Nelson}, P.~I. 1985, \apj, 293, 192

\bibitem[{{Ferland} {et~al.}(2013){Ferland}, {Porter}, {van Hoof}, {Williams},
  {Abel}, {Lykins}, {Shaw}, {Henney}, \& {Stancil}}]{2013RMxAA..49..137F}
{Ferland}, G.~J., {Porter}, R.~L., {van Hoof}, P.~A.~M., {et~al.} 2013, \rmxaa,
  49, 137

\bibitem[{{Foreman-Mackey} {et~al.}(2013){Foreman-Mackey}, {Hogg}, {Lang}, \&
  {Goodman}}]{2013PASP..125..306F}
{Foreman-Mackey}, D., {Hogg}, D.~W., {Lang}, D., \& {Goodman}, J. 2013, \pasp,
  125, 306

\bibitem[{Foreman-Mackey {et~al.}(2014)Foreman-Mackey, Price-Whelan, Ryan,
  Emily, Smith, Barbary, Hogg, \& Brewer}]{dan_foreman_mackey_2014_11020}
Foreman-Mackey, D., Price-Whelan, A., Ryan, G., {et~al.} 2014, triangle.py
  v0.1.1, doi:10.5281/zenodo.11020

\bibitem[{{Fox} {et~al.}(2005){Fox}, {Savage}, \&
  {Wakker}}]{2005AJ....130.2418F}
{Fox}, A.~J., {Savage}, B.~D., \& {Wakker}, B.~P. 2005, \aj, 130, 2418

\bibitem[{{Fumagalli} {et~al.}(2011{\natexlab{a}}){Fumagalli}, {O'Meara}, \&
  {Prochaska}}]{2011Sci...334.1245F}
{Fumagalli}, M., {O'Meara}, J.~M., \& {Prochaska}, J.~X. 2011{\natexlab{a}},
  Science, 334, 1245

\bibitem[{{Fumagalli} {et~al.}(2013){Fumagalli}, {O'Meara}, {Prochaska}, \&
  {Worseck}}]{2013ApJ...775...78F}
{Fumagalli}, M., {O'Meara}, J.~M., {Prochaska}, J.~X., \& {Worseck}, G. 2013,
  \apj, 775, 78

\bibitem[{{Fumagalli} {et~al.}(2011{\natexlab{b}}){Fumagalli}, {Prochaska},
  {Kasen}, {Dekel}, {Ceverino}, \& {Primack}}]{2011MNRAS.418.1796F}
{Fumagalli}, M., {Prochaska}, J.~X., {Kasen}, D., {et~al.} 2011{\natexlab{b}},
  \mnras, 418, 1796

\bibitem[{{Genel} {et~al.}(2014){Genel}, {Vogelsberger}, {Springel}, {Sijacki},
  {Nelson}, {Snyder}, {Rodriguez-Gomez}, {Torrey}, \&
  {Hernquist}}]{2014MNRAS.445..175G}
{Genel}, S., {Vogelsberger}, M., {Springel}, V., {et~al.} 2014, \mnras, 445,
  175

\bibitem[{{Haardt} \& {Madau}(2012)}]{2012ApJ...746..125H}
{Haardt}, F., \& {Madau}, P. 2012, \apj, 746, 125

\bibitem[{{Hinshaw} {et~al.}(2013){Hinshaw}, {Larson}, {Komatsu}, {Spergel},
  {Bennett}, {Dunkley}, {Nolta}, {Halpern}, {Hill}, {Odegard}, {Page}, {Smith},
  {Weiland}, {Gold}, {Jarosik}, {Kogut}, {Limon}, {Meyer}, {Tucker}, {Wollack},
  \& {Wright}}]{2013ApJS..208...19H}
{Hinshaw}, G., {Larson}, D., {Komatsu}, E., {et~al.} 2013, \apjs, 208, 19

\bibitem[{{Hopkins} {et~al.}(2007){Hopkins}, {Bundy}, {Hernquist}, \&
  {Ellis}}]{2007ApJ...659..976H}
{Hopkins}, P.~F., {Bundy}, K., {Hernquist}, L., \& {Ellis}, R.~S. 2007, \apj,
  659, 976

\bibitem[{{Isobe} \& {Feigelson}(1990)}]{1990BAAS...22..917I}
{Isobe}, T., \& {Feigelson}, E.~D. 1990, in Bulletin of the American
  Astronomical Society, Vol.~22, Bulletin of the American Astronomical Society,
  917--918

\bibitem[{{Jogee} {et~al.}(2008){Jogee}, {Miller}, {Penner}, {Bell},
  {Conselice}, {Skelton}, {Somerville}, {Rix}, {Barazza}, {Barden}, {Borch},
  {Beckwith}, {Caldwell}, {H{\"a}ussler}, {Heymans}, {Jahnke}, {McIntosh},
  {Meisenheimer}, {Papovich}, {Peng}, {Robaina}, {Sanchez}, {Wisotzki}, \&
  {Wolf}}]{2008ASPC..396..337J}
{Jogee}, S., {Miller}, S., {Penner}, K., {et~al.} 2008, in Astronomical Society
  of the Pacific Conference Series, Vol. 396, Formation and Evolution of Galaxy
  Disks, ed. J.~G. {Funes} \& E.~M. {Corsini}, 337

\bibitem[{{Kacprzak} {et~al.}(2012){Kacprzak}, {Churchill}, \&
  {Nielsen}}]{2012ApJ...760L...7K}
{Kacprzak}, G.~G., {Churchill}, C.~W., \& {Nielsen}, N.~M. 2012, \apjl, 760, L7

\bibitem[{{Kauffmann} {et~al.}(1993){Kauffmann}, {White}, \&
  {Guiderdoni}}]{1993MNRAS.264..201K}
{Kauffmann}, G., {White}, S.~D.~M., \& {Guiderdoni}, B. 1993, \mnras, 264, 201

\bibitem[{{Kere{\v s}} {et~al.}(2009){Kere{\v s}}, {Katz}, {Fardal},
  {Dav{\'e}}, \& {Weinberg}}]{2009MNRAS.395..160K}
{Kere{\v s}}, D., {Katz}, N., {Fardal}, M., {Dav{\'e}}, R., \& {Weinberg},
  D.~H. 2009, \mnras, 395, 160

\bibitem[{{Kimm} {et~al.}(2011){Kimm}, {Slyz}, {Devriendt}, \&
  {Pichon}}]{2011MNRAS.413L..51K}
{Kimm}, T., {Slyz}, A., {Devriendt}, J., \& {Pichon}, C. 2011, \mnras, 413, L51

\bibitem[{{Kriek} {et~al.}(2006){Kriek}, {van Dokkum}, {Franx}, {Quadri},
  {Gawiser}, {Herrera}, {Illingworth}, {Labb{\'e}}, {Lira}, {Marchesini},
  {Rix}, {Rudnick}, {Taylor}, {Toft}, {Urry}, \& {Wuyts}}]{2006ApJ...649L..71K}
{Kriek}, M., {van Dokkum}, P.~G., {Franx}, M., {et~al.} 2006, \apjl, 649, L71

\bibitem[{{Lavalley} {et~al.}(1992){Lavalley}, {Isobe}, \&
  {Feigelson}}]{1992BAAS...24..839L}
{Lavalley}, M.~P., {Isobe}, T., \& {Feigelson}, E.~D. 1992, in Bulletin of the
  American Astronomical Society, Vol.~24, Bulletin of the American Astronomical
  Society, 839--840

\bibitem[{{Lehner} {et~al.}(2013){Lehner}, {Howk}, {Tripp}, {Tumlinson},
  {Prochaska}, {O'Meara}, {Thom}, {Werk}, {Fox}, \&
  {Ribaudo}}]{2013ApJ...770..138L}
{Lehner}, N., {Howk}, J.~C., {Tripp}, T.~M., {et~al.} 2013, \apj, 770, 138

\bibitem[{{Levshakov} {et~al.}(2003){Levshakov}, {Agafonova}, {Centuri{\'o}n},
  \& {Molaro}}]{2003A&A...397..851L}
{Levshakov}, S.~A., {Agafonova}, I.~I., {Centuri{\'o}n}, M., \& {Molaro}, P.
  2003, \aap, 397, 851

\bibitem[{{Madau} \& {Haardt}(2009)}]{2009ApJ...693L.100M}
{Madau}, P., \& {Haardt}, F. 2009, \apjl, 693, L100

\bibitem[{{Marshall} {et~al.}(2008){Marshall}, {Burles}, {Thompson},
  {Shectman}, {Bigelow}, {Burley}, {Birk}, {Estrada}, {Jones}, {Smith},
  {Kowal}, {Castillo}, {Storts}, \& {Ortiz}}]{2008SPIE.7014E.169M}
{Marshall}, J.~L., {Burles}, S., {Thompson}, I.~B., {et~al.} 2008, in Society
  of Photo-Optical Instrumentation Engineers (SPIE) Conference Series, Vol.
  7014, Society of Photo-Optical Instrumentation Engineers (SPIE) Conference
  Series

\bibitem[{{Matejek} \& {Simcoe}(2012)}]{2012ApJ...761..112M}
{Matejek}, M.~S., \& {Simcoe}, R.~A. 2012, \apj, 761, 112

\bibitem[{{Mo} {et~al.}(1998){Mo}, {Mao}, \& {White}}]{1998MNRAS.295..319M}
{Mo}, H.~J., {Mao}, S., \& {White}, S.~D.~M. 1998, \mnras, 295, 319

\bibitem[{{Neistein} {et~al.}(2006){Neistein}, {van den Bosch}, \&
  {Dekel}}]{2006MNRAS.372..933N}
{Neistein}, E., {van den Bosch}, F.~C., \& {Dekel}, A. 2006, \mnras, 372, 933

\bibitem[{{Nelson} {et~al.}(2013){Nelson}, {Vogelsberger}, {Genel}, {Sijacki},
  {Kere{\v s}}, {Springel}, \& {Hernquist}}]{2013MNRAS.429.3353N}
{Nelson}, D., {Vogelsberger}, M., {Genel}, S., {et~al.} 2013, \mnras, 429, 3353

\bibitem[{{Prochaska} {et~al.}(2010){Prochaska}, {O'Meara}, \&
  {Worseck}}]{2010ApJ...718..392P}
{Prochaska}, J.~X., {O'Meara}, J.~M., \& {Worseck}, G. 2010, \apj, 718, 392

\bibitem[{{Rafelski} {et~al.}(2011){Rafelski}, {Wolfe}, \&
  {Chen}}]{2011ApJ...736...48R}
{Rafelski}, M., {Wolfe}, A.~M., \& {Chen}, H.-W. 2011, \apj, 736, 48

\bibitem[{{Rafelski} {et~al.}(2012){Rafelski}, {Wolfe}, {Prochaska},
  {Neeleman}, \& {Mendez}}]{2012ApJ...755...89R}
{Rafelski}, M., {Wolfe}, A.~M., {Prochaska}, J.~X., {Neeleman}, M., \&
  {Mendez}, A.~J. 2012, \apj, 755, 89

\bibitem[{{Rahmati} {et~al.}(2013){Rahmati}, {Pawlik}, {Rai{\v c}evi{\` c}}, \&
  {Schaye}}]{2013MNRAS.430.2427R}
{Rahmati}, A., {Pawlik}, A.~H., {Rai{\v c}evi{\` c}}, M., \& {Schaye}, J. 2013,
  \mnras, 430, 2427

\bibitem[{{Rauch} {et~al.}(1996){Rauch}, {Sargent}, {Womble}, \&
  {Barlow}}]{1996ApJ...467L...5R}
{Rauch}, M., {Sargent}, W.~L.~W., {Womble}, D.~S., \& {Barlow}, T.~A. 1996,
  \apjl, 467, L5

\bibitem[{{Sanders} {et~al.}(2015){Sanders}, {Shapley}, {Kriek}, {Reddy},
  {Freeman}, {Coil}, {Siana}, {Mobasher}, {Shivaei}, {Price}, \& {de
  Groot}}]{2015ApJ...799..138S}
{Sanders}, R.~L., {Shapley}, A.~E., {Kriek}, M., {et~al.} 2015, \apj, 799, 138

\bibitem[{{Savage} \& {Sembach}(1991)}]{1991ApJ...379..245S}
{Savage}, B.~D., \& {Sembach}, K.~R. 1991, \apj, 379, 245

\bibitem[{{Simcoe}(2011)}]{2011ApJ...738..159S}
{Simcoe}, R.~A. 2011, \apj, 738, 159

\bibitem[{{Simcoe} {et~al.}(2004){Simcoe}, {Sargent}, \&
  {Rauch}}]{2004ApJ...606...92S}
{Simcoe}, R.~A., {Sargent}, W.~L.~W., \& {Rauch}, M. 2004, \apj, 606, 92

\bibitem[{{Simcoe} {et~al.}(2008){Simcoe}, {Burgasser}, {Bernstein}, {Bigelow},
  {Fishner}, {Forrest}, {McMurtry}, {Pipher}, {Schechter}, \&
  {Smith}}]{2008SPIE.7014E..0US}
{Simcoe}, R.~A., {Burgasser}, A.~J., {Bernstein}, R.~A., {et~al.} 2008, in
  Society of Photo-Optical Instrumentation Engineers (SPIE) Conference Series,
  Vol. 7014, Society of Photo-Optical Instrumentation Engineers (SPIE)
  Conference Series, 0

\bibitem[{{Springel}(2010)}]{2010MNRAS.401..791S}
{Springel}, V. 2010, \mnras, 401, 791

\bibitem[{{Steidel} {et~al.}(2010){Steidel}, {Erb}, {Shapley}, {Pettini},
  {Reddy}, {Bogosavljevi{\'c}}, {Rudie}, \& {Rakic}}]{2010ApJ...717..289S}
{Steidel}, C.~C., {Erb}, D.~K., {Shapley}, A.~E., {et~al.} 2010, \apj, 717, 289

\bibitem[{{Toomre} \& {Toomre}(1972)}]{1972ApJ...178..623T}
{Toomre}, A., \& {Toomre}, J. 1972, \apj, 178, 623

\bibitem[{{Torrey} {et~al.}(2014){Torrey}, {Vogelsberger}, {Genel}, {Sijacki},
  {Springel}, \& {Hernquist}}]{2014MNRAS.438.1985T}
{Torrey}, P., {Vogelsberger}, M., {Genel}, S., {et~al.} 2014, \mnras, 438, 1985

\bibitem[{{van Dokkum} {et~al.}(2008){van Dokkum}, {Franx}, {Kriek}, {Holden},
  {Illingworth}, {Magee}, {Bouwens}, {Marchesini}, {Quadri}, {Rudnick},
  {Taylor}, \& {Toft}}]{2008ApJ...677L...5V}
{van Dokkum}, P.~G., {Franx}, M., {Kriek}, M., {et~al.} 2008, \apjl, 677, L5

\bibitem[{{van Dokkum} {et~al.}(2013){van Dokkum}, {Leja}, {Nelson}, {Patel},
  {Skelton}, {Momcheva}, {Brammer}, {Whitaker}, {Lundgren}, {Fumagalli},
  {Conroy}, {F{\"o}rster Schreiber}, {Franx}, {Kriek}, {Labb{\'e}},
  {Marchesini}, {Rix}, {van der Wel}, \& {Wuyts}}]{2013ApJ...771L..35V}
{van Dokkum}, P.~G., {Leja}, J., {Nelson}, E.~J., {et~al.} 2013, \apjl, 771,
  L35

\bibitem[{{Vogelsberger} {et~al.}(2014){Vogelsberger}, {Genel}, {Springel},
  {Torrey}, {Sijacki}, {Xu}, {Snyder}, {Nelson}, \&
  {Hernquist}}]{2014MNRAS.444.1518V}
{Vogelsberger}, M., {Genel}, S., {Springel}, V., {et~al.} 2014, \mnras, 444,
  1518

\bibitem[{{Werk} {et~al.}(2013){Werk}, {Prochaska}, {Thom}, {Tumlinson},
  {Tripp}, {O'Meara}, \& {Peeples}}]{2013ApJS..204...17W}
{Werk}, J.~K., {Prochaska}, J.~X., {Thom}, C., {et~al.} 2013, \apjs, 204, 17

\bibitem[{{Werk} {et~al.}(2014){Werk}, {Prochaska}, {Tumlinson}, {Peeples},
  {Tripp}, {Fox}, {Lehner}, {Thom}, {O'Meara}, {Ford}, {Bordoloi}, {Katz},
  {Tejos}, {Oppenheimer}, {Dav{\'e}}, \& {Weinberg}}]{2014ApJ...792....8W}
{Werk}, J.~K., {Prochaska}, J.~X., {Tumlinson}, J., {et~al.} 2014, \apj, 792, 8

\end{thebibliography}

\end{document}